\pdfoutput=1
\documentclass[a4paper,11pt,parskip=half-,abstract,bibliography=totocnumbered]{scrartcl}
\usepackage{geometry}
\usepackage{caption}
\DeclareCaptionLabelFormat{adja-page}{\hrulefill\\#1 #2 \emph{(previous page)}}

\usepackage[para]{threeparttable}
\usepackage{rotating} %
\usepackage{setspace} %
\usepackage{graphics}
\usepackage{amssymb,amsfonts,amsmath,amsbsy} %

\usepackage{enumitem}

\usepackage[longtable]{lineno} %

\usepackage{url}
\usepackage{array}
\usepackage{graphicx}
\usepackage{epstopdf}

\renewcaptionname{english}{\refname}{Bibliography} %
\usepackage{natbib}

\makeatletter
\newcommand\bibstyle@comma{\bibpunct(),a,,}
\newcommand\bibstyle@semicolon{\bibpunct();a,,}
\makeatother
\usepackage{etoolbox}
\pretocmd\citet{\citestyle{comma}}\relax\relax
\pretocmd\Citet{\citestyle{comma}}\relax\relax
\pretocmd\citep{\citestyle{semicolon}}\relax\relax
\pretocmd\Citep{\citestyle{semicolon}}\relax\relax
\pretocmd\citealp{\citestyle{semicolon}}\relax\relax
\pretocmd\Citealp{\citestyle{semicolon}}\relax\relax

\usepackage{nameref}
\usepackage{hyperref}
\hypersetup{
  colorlinks = true,
  citecolor=  black,
  linkcolor = blue,
  filecolor = cyan %
}
\usepackage{xurl} %

\newcommand*{\qpref}[1]{\hyperref[{#1}]{\textit{``\nameref*{#1}'', section \ref*{#1}}}}
\newcommand*{\qref}[1]{\hyperref[{#1}]{\textit{``\nameref*{#1}'' (section \ref*{#1})}}}

\newcommand*{\doi}{}
\makeatletter
\newcommand{\doi@}[1]{\href{https://doi.org/#1}{\textcolor{black}{DOI: } https://doi.org/#1}}
\DeclareRobustCommand{\doi}{\hyper@normalise\doi@}
\makeatother

\usepackage{pdflscape} %
\usepackage{authblk} %
\usepackage[iso,english]{isodate}

\usepackage[english]{babel}

\usepackage[utf8]{inputenc} %

\usepackage[]{newtx}
\usepackage[T1]{fontenc} %

\usepackage[export]{adjustbox}

\usepackage{zref-xr,zref-user}
\zxrsetup{toltxlabel}
\hypersetup{
  colorlinks = true,
  citecolor=  black,
  linkcolor = {blue},
  filecolor = cyan %
}

\usepackage[dvipsnames,table]{xcolor}

\newcommand{\blu}[1]{{\textcolor {blue} {#1}}}
\newcommand{\Burl}[1]{\blu{\url{#1}}}

\newcommand*{\FIXMO}[1]{}

\newcommand{\code}[1]{\textit{\textbf{Code:}} {#1}}

\newcommand{\citepos}[1]{\citeauthor{#1}'s \citeyearpar{#1}}
\newcommand{\citepospage}[2]{\citeauthor{#1}'s (\citeyear{#1}, #2)}
\usepackage[normalem]{ulem}

\usepackage{gitinfo2}
\usepackage[toc,page]{appendix}

\title{A structural causal framework for interventions on evolutionary accumulation models}

\author[1,2,*]{Ramon Diaz-Uriarte}
\author[1,2]{Íñigo Ríos Arroyo}
\author[3,4]{Iain G. Johnston}
\affil[1]{Instituto de Investigaciones Biomédicas Sols-Morreale (IIBM), UAM-CSIC,  Madrid, Spain}
\affil[2]{Department of Biochemistry, School of Medicine, Universidad Autónoma de Madrid, Madrid, Spain}
\affil[3]{Department of Mathematics, University of Bergen, Bergen, Norway}
\affil[4]{Computational Biology Unit, University of Bergen, Bergen, Norway}
\affil[*]{Author for correspondence: \texttt{r.diaz@uam.es}}

\cleanlookdateon
\begin{document}

\date{2026-06-27}

\maketitle

\clearpage

\begin{abstract}
  Evolutionary accumulation models (EvAMs), also known as cancer progression models (CPMs), infer dependencies in the order of accumulation of mutations during tumor progression from cross-sectional data. It has been suggested that EvAMs could be used to identify therapeutic targets, but there is no procedure in the literature for how to extract predictions under intervention from these models. A simple approach of conditioning on the absence of a mutation gives incorrect predictions. We address this gap by formalizing what ``intervene'' means for all currently available EvAM methods (OT, OncoBN, CBN, H-ESBCN, MHN, HyperHMM, HyperTraPS), using Pearl's \(do\) operator and conditional interventions. For each model, we show how to implement the intervention (in most cases as specific parameter modifications), identify equivalent implementation procedures, and analyze whether the modularity assumption ---required for the intervention to be well-defined--- is justified. Drawing on individual-level causal DAGs that make fitness an explicit variable, we distinguish two types of intervention (killing and inactivating) that are conflated in standard EvAM representations. Since the goal is to prioritize intervention candidates, we recast the problem as one of ranking: we define three intervention objectives and provide a protocol for evaluating how well EvAMs rank targets. Our framework is not specific to cancer or EvAMs; it applies wherever fitted computational models can be interpreted as structural causal models. Code available from \Burl{https://github.com/rdiaz02/scm-interv-evams}.

\end{abstract}

\section{Introduction}\label{sec:intro}

Cancer progression is driven largely by the sequential accumulation of somatic mutations during the life of an individual: cancer cells can accumulate driver mutations that increase cell-level fitness, increasing in frequency and spreading (and ultimately resulting in metastasis), at the expense of the whole organism \citep{Hanahan2011}.

Because of epistatic interactions, the accumulation of driver cancer mutations is often constrained to follow only some possible sequential orders, as the effect or viability of a mutation can depend on the presence of other mutations. Evolutionary accumulation models (EvAMs), also known as cancer progression models (CPMs), have been developed to infer these dependencies in the order of accumulation of mutations during tumor progression  from cross-sectional data (\citealp{Hainke2012, Beerenwinkel2015, diaz-uriarte2025}). From these dependencies, it is possible to derive the trajectories of tumor progression from an initial set of genotypes (e.g., wildtype) to the genotype with all driver mutations \citep{diaz-uriarte2019a, hosseini2019a, aga2024}.

EvAMs use cross-sectional data (data obtained from different samples which are taken at variable unknown time points ---\citealp{capri_bioinformatics, Beerenwinkel2015, diaz-uriarte2025}), a type of data which are very abundant, freely available for many cancer types, and easier to obtain than multiple samples from the same individual. These methods have also been used in very different domains (such as HIV mutation accumulation ---\citealp{beerenwinkel_evolution_2006, Beerenwinkel2007, montazeri_estimating_2015,posada-cespedes2021}---, and malaria progression in children ---\citealp{johnston2019}). This indicates their potential usefulness, and explains why new methods continue to be developed. Furthermore, EvAMs and related models could have a translational impact. Denoising data by inferring the most probable hidden genotype improves survival analysis of cancer data \citep{Gerstung2009}, and it has been shown that these methods can be used to stratify cancer patients \citep{angaroni2021,fontana2023} and could help triage malaria sufferers \citep{johnston2019}.

It has also been suggested \citep{Desper1999JCB, Attolini2010a, Gerstung2011} that EvAMs could be used to determine which genes  (or gene products or molecular pathways) should be targeted or intervened upon (the types of interventions considered are explained in section \ref{sec:what-mean-interv}). The simplest procedure would be to use these methods to identify initiating events, which are often more promising targets than late mutation acquisitions \citep{ozawa_most_2014, Cheng2012, cristea_pathtimex:_2017}. But the inferences of EvAM methods are  much richer than an ordering of early vs.\ late mutations, and it is not just order of accumulation, but particular consequences on downstream events that arguably matter most. For example, in so far as EvAMs can be used to predict evolutionary trajectories, we could use them to design interventions to drive the evolutionary process towards (or away from) chosen states \citep{hosseini2019a}.  Therefore, it should be possible to use EvAMs, and leverage on their direct and indirect inferences about the evolutionary process, to identify therapeutic targets for specific objectives.

From that perspective, instead of focusing on the quality of the inferences about the patterns of dependencies, what we want to know is what could work from the point of view of interventions. Thus, we shift the focus from ``correct'' to ``actionable'' inferences. It is conceivable that some inferences about the true dependencies can be wrong whereas ranking of genes for the purpose of interventions is successful and useful.
An EvAM will be useful to identify targets if it can provide a ranking of genes that is very similar (or highly correlated) to the true ranking of genes derived from the ground truth. But before this can be evaluated, we need a principled way to modify EvAMs to obtain predictions under intervention, a way to answer ``what if'' questions like ``What does this fitted model say would be the evolutionary dynamics if we were to intervene on gene \(A\)?''.  As we show in section \ref{sec:int-vs-cond}, treating intervention as conditioning on the absence of a mutation gives incorrect predictions.

Our focus is not the identification or estimation of causal effects from observational data ---the central concern of much of the causal inference literature \citep{hernan2020, morgan_counterfactuals_2015, pearl_causality_2009}. Rather, we address a logically prior question: given a fitted model, what is the correct procedure for extracting predictions under intervention? For models designed as regressions, this is straightforward: one substitutes the desired value into the fitted equation. But EvAMs were not designed for intervention, and the mapping from ``target gene \(A\)'' to a specific model manipulation is neither obvious nor uniform across model families.

We develop a rationale for modifying EvAMs that results from the
application of \citepos{pearl_causality_2009} \(do\) operator and conditional interventions \citep{correa2020a, eberhardt2007, korb2004} (both are formalisms for representing what would happen under an intervention, as distinct from observational conditioning). To draw the distinction between two types of intervention, killing vs.\ gene inactivation, that are conflated in standard EvAM representations, we build on the individual-level causal DAGs of \citet{frank97-price-causal} and \citet{otsuka_causal_2016, otsuka2019}, which make fitness an explicit variable in the causal structure.
We consider virtually all of the currently available EvAM methods, and for every intervention in each method, we include a detailed analysis of the role of modularity \citep{woodward_making_2003, pearl_causality_2009}.
We also discuss several ways of quantifying intervention effects and provide a protocol for evaluating EvAMs' ability to identify targets. Our notion of causal effects follows \citepospage{hernan2020}{p.~v} ``causal effects in populations, that is, numerical quantities that measure changes in the distribution of an outcome under different interventions.'' Since our goal is to prioritize intervention candidates rather than estimate individual effect sizes, we recast these quantities as a ranking problem (section \ref{sec:ranking-interv}).

The main contributions of our ms.\ are: (a) the conceptualization of what ``intervene'' means across a class of fitted models, using an interventionist causal framework, which provides a principled basis for manipulating the output of EvAMs to obtain predictions under intervention; (b) the differentiation between different types of intervention, which also sets the stage for possibly more complex schemes (like stochastic interventions); (c) the criteria to measure  gene importance from EvAMs.

\section{Overview}

To use EvAMs for interventions we need three things: (a) a clear definition of what an intervention is; (b) a principled way to modify fitted models to predict what happens under the intervention; (c) a way to quantify the effects of the intervention relative to an objective (the outcomes whose distribution changes under intervention). We address (a) below in subsection \ref{sec:what-mean-interv}, and (b), the key section of this ms., in section \ref{sec:cpm-modify-for-intervention}; (c) is discussed in  section \ref{sec:quant-interv-effects}. In the rest of this section we list the EvAMs considered (section \ref{sec:evams-considered}),  and provide a simple example of why a naive approach to intervention (conditioning) would fail with EvAMs (section \ref{sec:int-vs-cond}).

\subsection{EvAMs considered}\label{sec:evams-considered}

We consider here all EvAMs currently in use and with existing software implementations. These include Oncogenetic Trees (OT ---\citealp{Desper1999JCB, Szabo2008}), OncoBN \citep{nicol2021}, Conjunctive Bayesian Networks (CBN ---\citealp{Gerstung2009, montazeri_large-scale_2016}), Hidden Extended Suppes-Bayes Causal Networks (H-ESBCN ---\citealp{angaroni2021}), Mutual Hazard Networks (MHN ---\citealp{schill2020}), HyperHMM \citep{moen2023}, and HyperTraPS(-CT) \citep{johnston2016,greenbury2020,aga2024}. For overviews see \citet{diaz-uriarte2025, diaz-uriarte2022a, Beerenwinkel2015}.

\subsection{What do we mean by ``intervene on a gene''?}\label{sec:what-mean-interv}

We consider two different meanings of ``targeting a gene'': ``killing'' and ``inactivating'' interventions. In a killing intervention targeting a gene makes all genotypes that bear a mutation in that gene (i.e., a particular allele) lethal; equivalently, a killing intervention turns a mutation in that gene into a lethal mutation. This is coherent with widespread uses of this term and some of the existing  procedures in the field \citep{Luo2009, oneil_synthetic_2017-1}. A different type of intervention is the gene inactivating intervention \citep{dantonio2023,zhan2019} where, even if the gene is mutated, it has no downstream consequences, either because it is engineered back to the WT state or because its modified downstream products (e.g., mRNA, protein) are blocked; this is the type of intervention mentioned in \citet{schill2024b}.

\subsection{Intervention vs.\ conditioning}\label{sec:int-vs-cond}

\begin{figure}[tbh!]
   \centering \includegraphics[width=14.0cm,keepaspectratio]{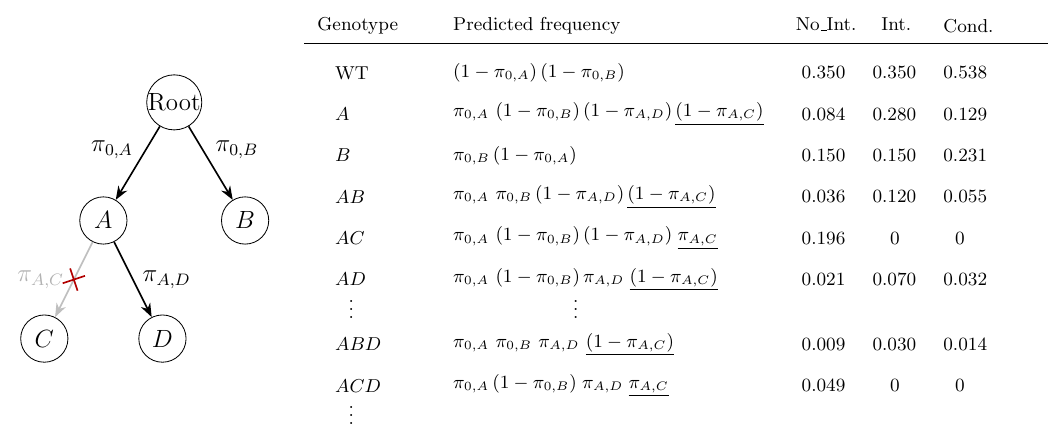}
   \caption{\textbf{Example intervention and predicted genotype frequencies.} On the left, an OT model. On the right, column ``Predicted frequency'' gives the predicted genotype frequencies of some of the genotypes (in the absence of errors: \(\epsilon_+ = \epsilon_- = 0\); see Appendix, section \ref{sec:genot_predictions_cpms_error}). Our intervention consists of making mutations in gene \(C\) lethal (i.e., \(C\) is the target). On the genotype frequency expressions, we underline the terms affected by the intervention, those that involve \(\pi_{A,C}\). As explained in the text, removing the edge leading to \(C\) (and any descendant edges from \(C\), if there were any) is equivalent to setting \(\pi_{A,C} = 0\). For example, the predicted frequency of genotypes \{A,C\} and \{A,C,D\} becomes 0. But the effect of the intervention is not the same as conditioning: setting to 0 the frequencies of all genotypes that have \(C\), and rescaling all genotypes to a total frequency of 1. For example,  the frequency of genotype \{B\} should not be affected by the intervention on \(C\), but  the frequency of genotype \{B\} would increase if we were to remove all genotypes with \(C\) and rescale to have frequencies sum up to 1. The effect of intervening on \(C\) (either inactivating \(C\) or making all genotypes with \(C\) mutated lethal ---section \ref{sec:what-mean-interv}) is that some \{A\} genotypes that would become \{A, C\} stay as \{A\} (the frequency of \{A\} increases as \(1 - \pi_{A,C}\) becomes 1), some  \{A, B\}s that would become \{A, B, C\} remain as \{A, B\}, etc. As a numerical example, columns ``No\_Int.'', ``Int.'' and ``Cond.'' correspond, respectively, to the genotype frequencies under no intervention, intervention on \(C\), and setting to 0 the frequencies of all genotypes that have \(C\), and rescaling,  for a model with \(\pi_{0, A} = 0.5, \pi_{0, B} = 0.3, \pi_{A, C} = 0.7, \pi_{A, D} = 0.2\). Similar examples can be constructed with all other models. }
\label{fig:ot_lethal_example}
\end{figure}

\citeauthor{pearl_causality_2009}'s \citeyearpar{pearl_causality_2009} \(do\) operator, written \(do(X = x)\), denotes intervention ---setting \(X\) to \(x\) by external action--- as distinct from conditioning on the observation \(X = x\). The two operations give different answers when \(X\) shares causes with, or is caused by, other variables in the model. For example, in a ``fork'' structure, where \(Z\) is a common cause of \(X\) and \(Y\) (\(Z \rightarrow X\), \(Z \rightarrow Y\)), observing \(X = x\) is informative about \(Z\) and thus about \(Y\), whereas intervening on \(X\) to set \(X = x\) makes \(X\) independent of \(Z\), and therefore independent of \(Y\),  so that \(P(Y|X = x)\) is generally different from \(P(Y|do(X = x))\). A separate issue arises with common effects (colliders): conditioning on a common effect of two variables induces a  dependence between them, biasing inference further down.

Now, let \(M_m\) denote the fitted model from EvAM method \(m\). We want to answer the question ``what would \(M_m\) predict if we were to intervene on gene \(g\)?''. A naive approach to predicting the consequences of interventions, removing genotypes with the targeted gene and rescaling, shown in Fig.~\ref{fig:ot_lethal_example}, is wrong. This operation would amount to conditioning, computing \(P(.|C = 0)\) (``among tumors without \(C\) mutated, what are the genotype frequencies?''), whereas we want \(P(. | do(C = 0))\) (``if we prevent \(C\) from mutating, what do genotype frequencies look like?''). The two distributions differ here because \(C\) shares causes with other events in the model; the collider \(A\) in Fig.~\ref{fig:cpm-modify-for-interventions}(d.2) shows where conditioning would additionally introduce collider bias.

\section{Modifying fitted EvAM models to predict the consequences of an intervention}\label{sec:cpm-modify-for-intervention}

To obtain the model predictions under intervention we construct \(M_{m,-g}\), the model for method \(m\) when we target gene \(g\).  Although we are dealing with different models, we modify them following a similar procedure that assumes modularity \citep{woodward_making_2003, pearl_causality_2009} and that takes advantage of interpreting these models as structural causal models; see Fig.~\ref{fig:cpm-modify-for-interventions}.
In all cases, the intervention is defined at the level of the variable (e.g., \(\sigma_A = do(A = 0)\)), meaning gene A is fixed as unmutated from \(t = 0\) onwards (the intervention is atomic but persistent: \(A\) is set to 0 at \(t = 0\) and held there throughout the evolutionary process); the specific implementations differ across model families.

Fig.~\ref{fig:cpm-modify-for-interventions} shows how each EvAM family represents an intervention. Panels (a)-(c) cover the individual-level causal DAG, where killing and inactivating are represented as different operations (conditional vs.\ atomic do). Panels (d)-(e) show the DAG of restrictions used by CBN, H-ESBCN, OT, and OncoBN, where the two interventions collapse to a single \(do\) intervention operation. Panels (f)-(g) show MHN, HyperHMM, and HyperTraPS, where the same collapse happens. The dashed arrows leading to panels (h)-(i) make the unifying point: across all these formalisms, the intervention yields the same kind of object ---a modified transition (rate) matrix.

The model-specific parameter changes (e.g., ``Set \(\lambda_A = 0\)'' in a CBN model) shown in Fig.~\ref{fig:cpm-modify-for-interventions}(e) and (g)  are the operational implementations of this single conceptual operation. The individual-level DAG interventions (Fig.~\ref{fig:cpm-modify-for-interventions}(b) and (c)) ---which apply only to CBN and H-ESBCN--- are equivalent to each other  at \(t = 0\) under strong selection weak mutation (SSWM), and thus to (e), as discussed in section \ref{sec:cbn-indiv-dag-interv}. In the rest of this section, we explain how to modify each model; algorithms in Appendix, section \ref{sec:appendix-interv-details}.

\begin{figure}[p!]
\centerline{\includegraphics[width=1.25\textwidth,keepaspectratio]{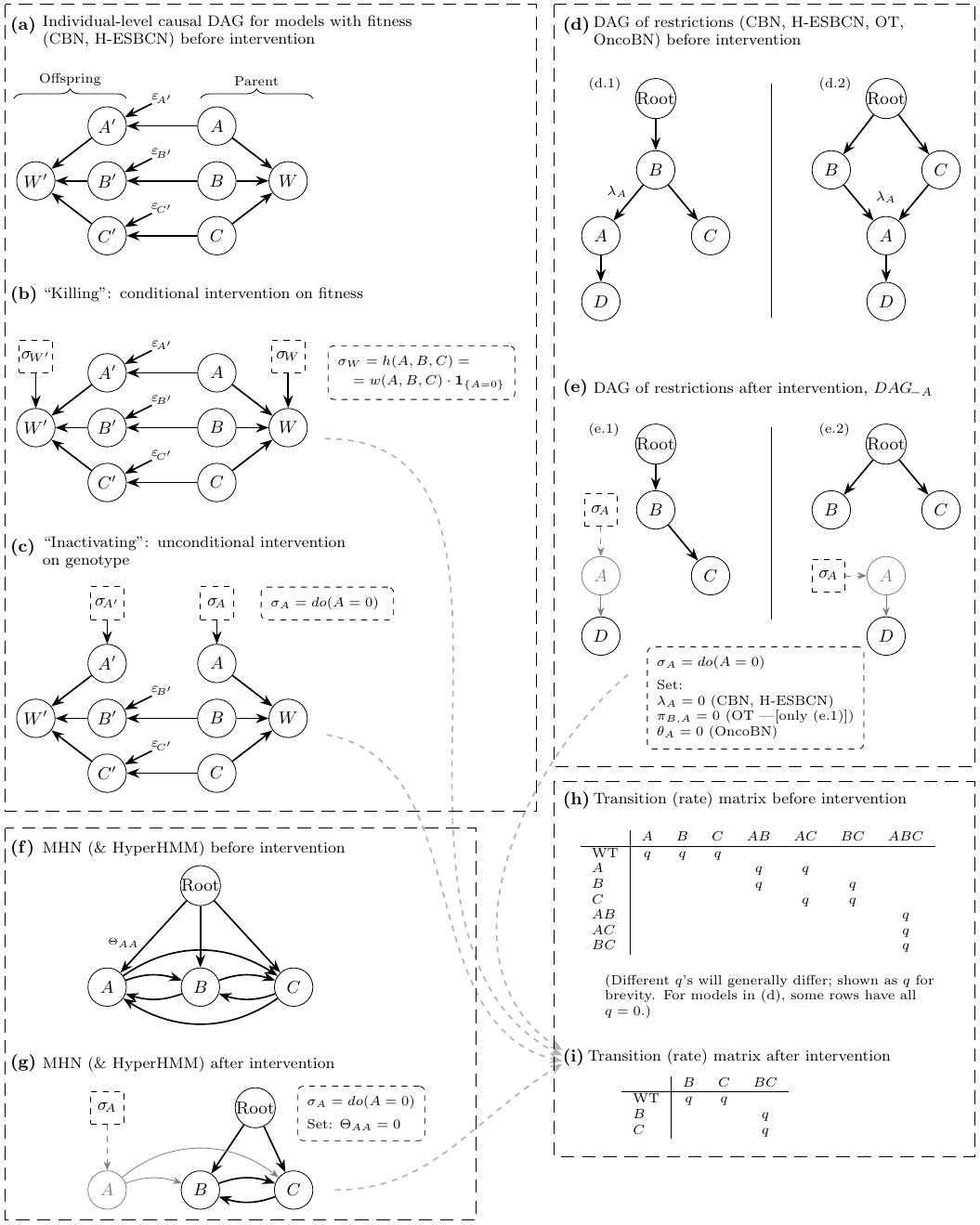}}
  \caption[]{\textit{(see next page)}}
\end{figure}
\begin{figure}[t!]
  \captionsetup{labelformat=adja-page,indention=-2.25cm,singlelinecheck=false}
  \ContinuedFloat
   \caption[Figure]{\textbf{Modifying EvAMs to predict an intervention.}
\textbf{(a) to (c)}. Individual level causal DAGs (following \citealp{frank97-price-causal}, Fig.~2; \citealp{otsuka2019}, Fig.~5.1; \citealp{edelaar2023}) for models with a fitness mapping (CBN, H-ESBCN). Fitness of both parent \(W\) and offspring (\(W'\)) is a function (given by the fitness landscape) of the status of three loci, \(A, B, C\). The status of the offspring's loci is a function of the parent's loci (plus a possible mutation).
\textbf{(b)} The ``killing'' intervention (make all genotypes that bear a mutation in a gene, here gene A, lethal) is a conditional intervention that leaves the original causal structure intact but (conditionally) replaces the fitness value; denoted as \(\sigma_{W}\) following \cite{correa2020a} (notation identical to their Example 1, using \(h(.)\) instead of \(g(.)\),  \(W\) instead of \(X\), and \(A, B, C\) instead of \(w\)).
\textbf{(c)} Inactivating a gene is an unconditional intervention on the genotype, denoted with the \(do\) operator \citep{pearl_causality_2009}; intervention in the offspring, \(\sigma_{A'}\), ``mutilates'' the DAG by removing the parental (and error) incoming arrows.
\textbf{(d)} DAGs of restrictions (CBN, H-ESBCN, OT, OncoBN); (d.2) cannot be represented by OT (two parents for \(A\)).
\textbf{(e)} Intervention on the DAGs of restrictions. This is an unconditional atomic \(do\) intervention that also mutilates the DAG of restrictions; here we cannot differentiate between killing and inactivating (see text). We can implement this intervention by setting a model parameter to 0. This intervention yields \(DAG_{-g}\) in the text.
\textbf{(f)} MHN and HyperHMM model before intervention.
\textbf{(g)} MHN after intervention. As in (e) we cannot differentiate between inactivating and killing. The \(do\) operation mutilates the cyclic graph (removing all incoming arrows into \(A\)). Implemented for MHN by setting \(\Theta_{A,A} = 0\) (equivalently, \(\theta_{A,A} = -\infty\)); for HyperHMM and HyperTraPS it involves modifying the conditional transition matrix ---see text.\\
\textit{Notation:} \(A, B, C, D\): loci; each locus can be in the wildtype (0) or mutated (1) states. \(W\): fitness. \(\varepsilon_i'\): transmission error term (i.e., mutation) for locus \(i\). \(w(A, B, C)\): fitness mapping function which gives the value of fitness of the genotype with specific states of loci A, B, C (i.e., the value of \(W\) corresponding to a particular genotype in the fitness landscape).
\(\boldsymbol{1}_{\{A = 0\}}\): indicator function that takes value 1 when gene A is not mutated, and 0 otherwise.
\(\lambda_A, \pi_{B,A}, \theta_A\): parameters in the DAGs of restrictions from CBN (and H-ESBCN), OT, and OncoBN, respectively (only \(\lambda_A\) shown annotating the DAG; \(\pi_{B,A}, \theta_A\) would annotate the same edge). \(\Theta_{A,A}\): MHN rate of event \(A\) before other events are present.
\(q\): parameters of the transition (rate) matrix from one genotype to another; different \(q\) will often be different.\\
We show four loci in (d) and (e) to illustrate the consequences of the intervention and DAG mutilation on nodes that descend from the one affected (i.e., only the incoming edges into the affected gene are affected; outgoing edges and parameters remain unchanged ---see text for details).
We only show three genes in the remaining panels to minimize clutter.
Panels (a), (b), (c), only show genotype and fitness (like Fig.~2 in \citealp{frank97-price-causal}), as that is sufficient for our models.  Individual-level causal DAGs can, however, include phenotype  (e.g., Fig.~3 in \citealp{frank97-price-causal}, Fig.~5.1 in \citealp{otsuka2019}) and if the inactivating intervention (panel c) acted on messenger RNA, proteins, etc, other mediating variables as needed.
The modified models in (b) (and (c)), (e), and (g) correspond to \(M_{m,-g}\) (with \(g = A\)), in step \ref{EvAM_I_g} of  \qref{sec:cpm-interv-pred}. For HyperHMM and HyperTraPS, see sections \ref{sec:hyperhmm-interv} and \ref{sec:hypertps-interv}. Algorithms for the different interventions in Appendix, section \ref{sec:appendix-interv-details}.
   }
\label{fig:cpm-modify-for-interventions}
\end{figure}

\subsection{CBN and H-ESBCN: individual-level causal DAG interventions}\label{sec:cbn-indiv-dag-interv}

Under the SSWM evolutionary regime, where the population is monomorphic except during brief clonal sweeps and evolution proceeds only uphill in fitness (Appendix, \ref{sec:fl-to-trm}), the CBN and H-ESBCN models implicitly specify a fitness landscape (``(...) the assignment of fitness values to genotypes that are connected by mutations'' ---\citealp[p.~2]{krug2021}) whose transition rate matrix is identical (up to a scaling constant) to that of the original EvAM model (Appendix, \qpref{sec:dags-to-fitness}).

This allows us to define interventions directly on the fitness landscape. Intervention by killing means setting the fitness of the affected genotypes (those bearing a mutation in the intervened upon gene) to 0; modularity requires that the fitness of all other genotypes remains unaffected.

In Fig.~\ref{fig:cpm-modify-for-interventions}(a), the causal DAG shows the individual-level causal relationships between genotype of the parent, genotype of the offspring, and fitness,  following \cite{frank97-price-causal}, \cite{otsuka_causal_2016, otsuka2019}, \cite{edelaar2023}. On this DAG, intervention by killing is not an atomic \(do\) intervention, but a conditional (\citealp{correa2020a}; \citealp[section 4.2]{pearl_causality_2009}), soft or parametric \citep{eberhardt2007}, or dependent (\citealp{korb2004}; \citealp[section 3.8.3]{korb_bayesian_2011}) intervention (Fig.~\ref{fig:cpm-modify-for-interventions}(b)): the original fitness mapping function, \(w(.)\) is replaced by a new function, \(h(.)\),
implementing the modification described above, while the original causal structure remains intact (there is no DAG mutilation). The population-level dynamics then follow from the continuous-time Markov chain under SSWM, where the transition rates between genotypes are determined by the fitness landscape (Appendix, \qref{sec:fl-to-trm}).

In addition to ``killing'', it is worth considering a different intervention (section \ref{sec:what-mean-interv}), shown in Fig.~\ref{fig:cpm-modify-for-interventions}(c). This intervention involves ``inactivating'' a gene (or its products), and is the intervention mentioned in \cite[Fig.~3]{schill2024b} (``A drug treatment that suppresses event 2'').  It is not a conditional intervention, but the usual atomic \(do\) intervention. For example, inactivation of the mutant allele of gene \(A\) makes the phenotype of any genotype with \(A\) mutated identical to the phenotype of that same genotype without \(A\) mutated. Mutations in \(A\) are turned into neutral mutations (that could still fix by hitchhiking or drift ---though not by drift under our strict SSWM model where we only move uphill in fitness) and no longer have any downstream consequences.

  If the interventions are carried out at \(t = 0\), before the evolutionary process starts, both killing and inactivating a gene are equivalent (compared to inactivating, killing leads to an increase in death rate because mutations that inactivation renders neutral lead to death under killing, but this difference is negligible for sensible values of mutation rate ---recall we assume SSWM and thus \(N \mu \ll 1\): Appendix, section \ref{sec:fl-to-trm}). Both inactivation and killing halve the size of the genotype space.  Focusing on phenotypes, under both interventions the observable distinct states are the same, and the transition rates between phenotypes are the same  (if the targeted gene is \(A\), the population evolves in the sub-hypercube with \(A = 0\)). But killing and inactivating won’t necessarily be identical if we intervene later. A simple example: populations made entirely of the \(\{A, B\}\) genotype, under killing, will disappear; however, under inactivation of \(A\) those populations do not disappear, but now look like a population of \(\{B\}\)s, that could then transition to \(\{B, C\}\).

  What about the implementation of the inactivating intervention? When this intervention is carried out at \(t = 0\), we will never observe any genotype with \(g\) mutated. Since these genotypes are now no longer part of the state space, this is the same as removing them from the transition rate matrix derived from the fitness landscape.  Under SSWM, this transition rate matrix is, therefore, identical to the one obtained by setting the fitness of the affected genotypes to 0. (If the intervention were carried out at a later \(t\), a more complex algorithm would be required. Suppose the transition rate matrix is of the form \(Q_{from, to}\), with rows as origin and columns as destination. Then, genotypes with \(g\) mutated will have a 0 as destinations ---i.e., all entries in columns corresponding to genotypes with \(g\) mutated would be 0. As for rows, the row of every genotype \(y\) with \(g\) mutated mirrors the row of \(x = y \setminus \{g\}\): the rate of \(y\) gaining mutation \(i \neq g\) is set to the original rate of \(x\) gaining \(i\)).

  \subsection{CBN and H-ESBCN: DAG of restrictions interventions}\label{sec:cbn-dag-restr-interv}

Independently of the individual-level causal DAG, CBN and H-ESBCN are structural causal models. The CBN model is also known as the noisy-AND causal model \citep{Beerenwinkel2007}. For example, in Fig.~\ref{fig:cpm-modify-for-interventions}(d.2), the structural equation for the waiting time to event \(A\) is given by \(T_A = \max(T_B, T_C) + U_A\), where \(U_A \sim \operatorname{Exp}(\lambda_A)\).  A similar reasoning applies to H-ESBCN, where the waiting times to events are also given by structural equations (with appropriate adjustment for OR/XOR relationships).

For CBN and H-ESBCN, from the DAG of restrictions that specifies the model, Fig.~\ref{fig:cpm-modify-for-interventions}(d), we use an atomic \(do\) intervention (which also involves DAG mutilation) fixing the gene as not-mutated. We can achieve this by setting \(\lambda_A = 0\) (so \(T_A = \infty\)) thus preventing  \(A\) from ever gaining a mutation. Because  \(\lambda_i\) are mechanism parameters that specify the rate at which \(i\) happens (given its restrictions are met) modularity holds: we can intervene on one node, without affecting the remaining structural equations. (This also explains why, in  Fig.~\ref{fig:cpm-modify-for-interventions}(e), we do not modify the \(A \rightarrow D\) relationship, and \(do(A=0)\) will lead to \(D\) never arising).

Setting \(\lambda_g = 0\) in the model, \(DAG_{\lambda_g = 0}\), has the same consequences (in terms of genotypes that can exist and rates of transitions between genotypes) as eliminating,
from the DAG that specifies the model, the node of the targeted gene and any other descendant nodes that strictly depend on the targeted gene
(see further details and algorithm in Appendix, \qref{sec:cpm_interv_dag}).
We denote this modified DAG of restrictions as \(DAG_{-g}\). (And this is the default procedure we use in the code.)
In addition, a third equivalent procedure is removing, from the fitted transition rate matrix \(Q\), all genotypes with gene \(g\) mutated, (\(Q_{-g}\)). Thus, \(DAG_{\lambda_g = 0}\), \(DAG_{-g}\), and \(Q_{-g}\) are equivalent ways of obtaining \(M_{m, -g}\).

  Those three procedures result in the same predictions as modifications of the fitness landscape (see Appendix, \qref{sec:appendix-interv-flandscape}), as we would expect from the structural causal model interpretation. This illustrates the expressive power of conceptualizing the interventions using individual-level causal DAGs, as well as the consistency of interventions guided by the modularity assumption across formalisms: the individual-level DAG (Fig.~\ref{fig:cpm-modify-for-interventions}(a)-(c)) and the DAG of restrictions (Fig.~\ref{fig:cpm-modify-for-interventions}(d)-(e)).

  However, we cannot differentiate between ``killing'' and ``inactivation'' on the DAG of restrictions, in contrast to the individual-level causal DAGs. This is a consequence of the different granularity of the models in Fig.~\ref{fig:cpm-modify-for-interventions}(a) and (d). Under (a) the model has the structure \(Genotype \rightarrow Fitness \rightarrow Evolutionary\ Dynamics\); because both \(Fitness\) and \(Genotype\) are intermediate variables, we can intervene on each one separately, (killing: conditionally change fitness; inactivation: unconditionally change genotype or phenotype) and they remain distinguishable. In contrast, the DAG of restrictions (and MHN, HyperHMM, and HyperTraPS models) have the structure \(Genotype \rightarrow Evolutionary\ Dynamics\). Fitness is, if at all, only implicit in the transition rate matrix, and the model structure is not expressive enough to differentiate the two interventions. There is no ambiguity if the intervention is performed before the process starts. But if the intervention is performed after the evolutionary process starts, we would need to specify which kind of intervention we are conducting. We would then need to modify not only the DAG, but also the transition rate of the existing populations: removing rows (killing) or transforming them (inactivating).

\subsection{OT and OncoBN: DAG of restrictions interventions}\label{sec:ot-dag-restr-interv}

OT and OncoBN also specify restrictions as DAGs, but they are untimed models. OT and OncoBN give the probability of observing a particular genotype at the time of sampling under some sampling scheme, so their parameters conflate the causal mechanism (e.g., the DAG + \(\lambda\)s of CBN and H-ESBCN) with the sampling process.
In contrast, in CBN and H-ESBCN the \(\lambda_i\)s are invariant under changes in the sampling distribution: from the continuous-time Markov chain, we can obtain the probability of observing any genotype under arbitrary sampling distributions. And under the interventionist framework \citep{pearl_causality_2009, woodward_making_2003}, we require that under DAG mutilation the remaining structural equations stay invariant. With CBN and H-ESBCN modularity holds provided the \(\lambda_i\)s describe independent mechanisms. But this is harder to justify under OT/OncoBN, which seem context (sampling process) dependent. For example, if the sampling distribution were to change (say, from exponential to uniform), the \(\pi\) parameters of OT would change even though the underlying causal mechanism has not. Still, this intervention where we hold the remaining parameters fixed is all we can do for the OT/OncoBN models without invoking additional assumptions.

As with CBN and H-ESBCN, we also denote the modified DAG as \(DAG_{-g}\). Also as for CBN and H-ESBCN, setting \(\pi_{parent(g), g} = 0\) (OT) or \(\theta_g= 0\) (OncoBN) has the same consequences as eliminating, from the DAG that specifies the model, the node of the targeted gene, \(g\), and any other descendant nodes that strictly depend on \(g\) (see algorithm in Appendix, \qref{sec:cpm_interv_dag} ---as for CBN and H-ESBCN, and for implementation reasons, \(DAG_{-g}\) is the default procedure in the code). In contrast to CBN and H-ESBCN, however, directly modifying fitness (as in Fig.~\ref{fig:cpm-modify-for-interventions}(a)-(c)) is not possible with OT and OncoBN because of a lack of equivalence between these models and fitness landscapes \citep{diaz-uriarte2025}.

\subsection{MHN: graph interventions}
\label{sec:mhn_graph-interv}

Fig.~\ref{fig:cpm-modify-for-interventions}(f) shows an MHN model as a graph. This representation differs from \cite[Fig.~1(g) in][]{schill2020} by adding the ``Root'' node, similar to Fig.~\ref{fig:cpm-modify-for-interventions}(d). We make the Root node explicit to highlight the similarity with the other models and emphasize that the \(\Theta_{i,i}\) are analogous to the \(Root \rightarrow i\) in the DAGs of restrictions models. This is also a structural causal model but the model for genes, in contrast to Fig.~\ref{fig:cpm-modify-for-interventions}(d), includes cycles: gaining event \(A\) has an effect on the probability of gaining \(B\), but gaining \(B\) also affects the probability of gaining \(A\).
Cycles can introduce several complications, including in the semantics of interventions \citep{bongers2021}. However, the MHN process is irreversible (mutations are gained, not lost) and sequential: for a given genotype, the transition rates to other genotypes are well-defined and there is no simultaneity problem. And the MHN model also specifies structural equations: the rate at which gene \(A\) is acquired is a deterministic function of the current mutation state of all other genes (\(q_{x\rightarrow x+A} = \Theta_{AA} \prod_{x_j=1} \Theta_{Aj}\) --- \citealp[eq.~2]{schill2020}), and thus we can express the time to event \(A\), given the state of the other genes, as a deterministic function plus an exogenous noise term (e.g., \(T_A = f_A(x, \varepsilon_A)\), where \(x\) is the current state of the genotype, \(\varepsilon_A\) a uniform \((0, 1)\) noise term, and \(f_A\) the inverse cumulative distribution function of an exponential with parameter \(q_{x\rightarrow x+A}\) ).

Thus, as for the DAGs of restrictions for CBN and H-ESBCN (section \ref{sec:cbn-dag-restr-interv}; Fig.~\ref{fig:cpm-modify-for-interventions}(d)-(e)), we intervene on the \(A\) variable with an atomic \(do(A = 0)\); we implement this intervention by setting \(\Theta_{AA} = 0\) (or, equivalently, \(\theta_{AA} = - \infty\)). This amounts to eliminating \(A\) from the \(\Theta\) matrix, yielding \(\Theta_{-A}\), as no genotype with that gene mutated is possible. Finally, we can also remove, from the fitted transition rate matrix \(Q\), all genotypes with gene \(A\) mutated, (\(Q_{-A}\)).

For simplicity, Fig.~\ref{fig:cpm-modify-for-interventions}(f) shows the original model of \citet{schill2020}. But the reasoning above applies, without modification, to the extended MHN model in \citet{schill2024b} that corrects for collider bias: preventing \(A\) from gaining a mutation ---the \(do(A = 0)\) intervention--- propagates downstream (to effects on the observation rate or on the rates of acquisition of other events) as per modularity.

As for CBN and H-ESBCN, \(\Theta_{g,g} = 0\), \(\Theta_{-g}\), and \(Q_{-g}\) are equivalent ways of implementing the modified model, \(M_{m, -g}\). The modularity assumption in this case plays a similar role as for CBN and H-ESBCN: in so far as the \(\Theta_{ij}\) describe independent mechanisms, when we intervene by preventing \(A\) from being gained, we do not affect the remaining structural equations. The parameters of MHN cannot be interpreted in terms of fitness landscapes \citep{diaz-uriarte2025}; still, MHN's parameters are directly interpretable in terms of rates and effects and invariant under changes in the sampling distribution (in contrast to OT and OncoBN). Therefore, the modularity assumption seems more plausible for MHN than OT and OncoBN,  though not as well supported as for CBN and H-ESBCN when these two can be mapped to fitness landscapes.

\subsection{HyperHMM: interventions via the conditional transition matrix}
\label{sec:hyperhmm-interv}

HyperHMM, similar to MHN, defines, for each genotype \(x\), the probability of gaining a new mutation. The (implicit) structural equation would be similar, but instead of events having pairwise interactions, interactions can be of arbitrary order (i.e., gaining \(A\) might be affected not just by the independent effects of the state of \(B\) on gaining \(A\) and the effect of \(C\) on gaining \(A\), but by the effect of the \(B, C\) state on gaining \(A\)). Thus, Fig.~\ref{fig:cpm-modify-for-interventions}(f) can also describe HyperHMM. However, HyperHMM returns a conditional transition matrix, where each row specifies the probability of the next genotype, given that there is a transition; the normalization couples all transition probabilities from a given state. When we intervene on \(A\), the probability mass assigned to \(A\) has to go somewhere. Removing \(A\) and renormalizing (e.g., if considering gaining \(B\), \( P(x \rightarrow x + B | \text{transition, no A})  = P(x \rightarrow x + B) / (1 - P(x \rightarrow x + A)) \)) would violate modularity because we change the mechanism for gaining \(B\). Alternatively, we can absorb that probability into a self-transition: the probability mass that would go to \(A\) keeps the population at state \(x\), while absolute rates for other transitions are unchanged. This preserves modularity: if HyperHMM's conditional probabilities are viewed as arising from underlying rates (with these rates being the ``true'' structural parameters), then removing \(A\) means removing one rate while keeping the others fixed. This is what the self-transition adjustment achieves.

In more detail, from the HyperHMM conditional transition matrix, we want to predict population-level genotype frequencies when some genotypes are now lethal (i.e., when transitions to lethal genotypes are impossible). Each step in a discrete-time process that we index by the number of mutations can be seen as an “attempt at mutation.” With lethal genotypes, some attempts fail to generate a viable transition and so the population remains in its current state.
Suppose we intervene on gene \(A\), i.e., we make the mutant allele of gene \(A\) lethal. We are at genotype \(x\), that does not contain \(A\) mutated; \(x+A\), genotype \(x\) plus \(A\) mutated, is now lethal.
  Under the SSWM regime used here, with small mutation rate, the limiting factor for (population-level) transitions between genotypes is mutation rate. That \(x\rightarrow x + A\) is now lethal does not affect the (per-unit-time) mutation rate to other non-lethal alleles  (i.e., for any \(y\) a non-lethal genotype, the mutation supply to \(y\) is unaffected). Thus, the absolute rates (or the probability per unit time) of the population moving from \(x\) to \(y\) (where \(y\) is non-lethal) is unchanged. (Though conditionally, if a transition occurs, it is now relatively more likely to a non-lethal genotype.)
  Therefore, since the population cannot transition to genotype \(x+A\), at the next attempt at mutation, all the population-level flux that went from \(x\) to \(x+A\) stays at \(x\); in other words, all ``attempts at mutation'' that involve \(A\) fail to generate a viable transition, so the population remains in its current state.
  In contrast, the flux from \(x\) to anything not \(x+A\) is not altered.

Based on that argument, we use the following procedure. Let \(R\) be the original conditional transition matrix. Suppose we intervene on gene \(g\). For a genotype \(x\) that does not contain \(g\) mutated, \(r_{x \rightarrow x+g}\) is the probability  of a transition from genotype \(x\) to the now-lethal combination, \(x+g\).

\begin{enumerate}
\item For every genotype \(x\) (that does not contain \(g\) mutated), set \(r_{x\rightarrow x}\) to \(r_{x \rightarrow x+g}\), and set \(r_{x \rightarrow x+g}\) to 0 (i.e., genotypes that contain gene \(g\) cannot be transitioned to, equivalent to having their rows zeroed.) This yields \(R_{-g}\) (the HyperHMM implementation of \(M_{m, -g}\)).

\item With \(R_{-g}\) obtain \(\hat{\mathbf{h}}_{-g}\), the predicted hitting probabilities after intervening in gene \(g\).

\item With \(R_{-g}\) obtain \(\hat{\mathbf{f}}^v_{-g}\), the predicted distribution of genotypes after \(v\) steps (\(v\) taking all integer values from 0 to the number of loci), as per the usual \(\hat{\mathbf{f}}^v_{-g} = \mathbf{f}^0\ R_{-g}^v\), where \(\mathbf{f}^0\) is the (row) vector \((1, 0, 0, ....)\).

\item Obtain the predicted population composition via the weighted sum \\ \(\hat{\mathbf{f}}_{-g} = \sum_{v=0}^{v=\text{number of loci}}\hat{\mathbf{f}}^v_{-g}\  P(V = v)\). \(P(V=v) \) is the empirical (observed) frequency, in the training sample, of observations with \(v\)  mutations. Equivalently (since in the original, unintervened data, each step produces exactly one mutation), \(P(V=v) \) is the frequency of samples that have undergone \(v\) steps. Also because of this equivalence, the summation goes from 0 to the number of loci, \(L\), which is the maximum number of steps any sample in the original data can have undergone.

    Post-intervention, \(v\) no longer equals the number of mutations (some steps are self-transitions), but we keep the summation bounded by \(L\): \(P(V=v)\) is empirically defined from the unintervened training data, which has \(v \leq L\); extending the sum beyond \(L\) would require extrapolating \(P(V=v)\) for \(v > L\), for which there is no empirical grounding.

\end{enumerate}

  With HyperHMM, the interpretability of interventions is weaker than with MHN: we need to construct the self-transition adjustment, which requires the additional argument about rates vs.\ conditional probabilities; thus, the assumption of modularity crucially hinges on the robustness of these adjustments. The mechanistic interpretability of this intervention, and the appeal to modularity, however, is arguably better grounded for HyperHMM than for OT and OncoBN: HyperHMM's conditional probabilities are not sampling-regime dependent in the way that OT and OncoBN's parameters are.

\subsection{HyperTraPS}\label{sec:hypertps-interv}

HyperTraPS, when used with cross-sectional data, returns as output a transition matrix on the hypercube structurally identical to HyperHMM's. This is the case regardless of the model structure \citep{aga2024} used:  zero-parameter, \(L\)-parameter (independent, non-interacting features), \(L^2\) or pairwise interactions (similar to MHN), \(L^3\) (three-way interactions), \(L^4\) (four-way interactions), and independent rates on each edge of the hypercube (like HyperHMM).
Therefore, to intervene on gene \(g\), and regardless of model structure or whether regularisation has been used during fitting, we implement \(do(g = 0)\) using the exact same procedure as for HyperHMM, section \ref{sec:hyperhmm-interv}, with \(R\) replaced by HyperTraPS's output transition matrix.
(Even in models where transitions are expressed in terms of individual parameters ---from independent to four-way---, zeroing parameters, as we do for MHN, would not correctly implement \(do(g = 0)\): the probability mass of the transitions to the now-lethal genotypes would not be assigned to a self-transition.)

\section{Quantifying intervention effects: intervention objectives}
\label{sec:quant-interv-effects}

Section \ref{sec:cpm-modify-for-intervention} discusses how to modify the models to obtain predictions under interventions. To properly assess if EvAMs can be used to identify therapeutic targets, we also need to specify what ``identify'' means, as different objectives can lead to different rankings of the same genes as targets.

Here we propose three intervention objectives that reflect three different ways of quantifying intervention effects and can be used in studies that evaluate the usefulness of EvAMs.
The ground truth of these intervention objectives can be measured from empirical data (e.g., from clinical experimental studies) or from simulated data (e.g., by rerunning evolution on fitness landscapes where we modify the fitness of specified genotypes).
The predicted values come from the EvAM methods themselves, independently of any fitness landscape interpretation: an EvAM yields hitting probabilities or genotype frequencies after intervention via the method-specific manipulations described in section \ref{sec:cpm-modify-for-intervention}, and these are what we compare against the ground truth.
We can then assess the usefulness of these approaches with respect to a given objective: an EvAM will be useful to identify targets if it can provide a ranking of genes, with respect to the specified objective, that is very similar (or highly correlated) to the true ranking of genes defined from the ground truth. We first describe the objectives, and next explain how to evaluate the performance of EvAMs in simulation studies; a protocol is provided in the Appendix \qref{sec:assess-interv-object}.

\begin{description}[style=nextline]

\item[Increase/decrease the probability of going through (hitting) a given genotype\\ (O\_genotype):]  Minimize or maximize the probability that evolution, starting from the WT genotype, will ever go through, or hit, a specific genotype; for example, minimize for genotypes related to metastasis, and maximize for genotypes that can be easily targeted by other means.
The simplest and most effective  way to avoid a given genotype would be to make lethal any of the genes that it has mutated, but we do not consider this a suitable intervention, because it is trivial. Instead, we can focus on interventions on genes, different from the ones mutated in the genotype of interest,  that can cause changes in the hitting probability of the genotype of interest by operating via the causal dependencies in the acquisition of genes.

\item[Decrease overall frequency of mutations or alterations (O\_mut):]
  Make the average number of mutations as small as possible.

\item[Increase frequency of the non-mutated, starting genotype, WT (O\_WT):]
  Delay the accumulation of altered genotypes, regardless of the number of mutations.
\end{description}

The first objective, O\_genotype, involves steering evolution in specific directions (to either avoid, or go through, particular stages). %
O\_genotype is independent of sampling time (it is the hitting, or first passage, probability of the underlying Markov chain). O\_WT and O\_mut are unavoidably dependent on sampling time, since both are obtained from the genotype frequencies at a particular time (section \ref{sec:pred-genot-freqs-time}).  Details on how to assess these three intervention objectives are provided in the Appendix, \qref{sec:assess-interv-object}.
We are not considering measures related to intra-tumor heterogeneity or diversity. The EvAMs considered here cannot be used to provide these predictions, since they are either models of between-subject bulk samples or between-subject single cell samples \citep{diaz-uriarte2025}; furthermore, in this paper we are assuming SSWM, so by assumption within-tumor heterogeneity is 0 (except for brief periods of fast clonal sweep).

\subsection{Obtaining the ground truth of intervention effects in simulation studies}\label{sec:ground-predictions}

If we assume SSWM, for any fitness landscape we can obtain the transition rate matrix \(\mathcal{Q}\) under SSWM (Appendix, \qpref{sec:fl-to-trm}). From \(\mathcal{Q}\), and via competing exponentials, we can obtain the embedded discrete-time chain and from its transition matrix the hitting (or first passage) probabilities (\citealp{norris1997, privault2018}), \(\mathbf{h}\), starting from the WT state. From \(\mathcal{Q}\) we can also obtain the distribution of genotypes, \(\mathbf{f}\), at any specified time. Assuming SSWM does not require us to focus on additional complicating factors such as bulk sequencing \citep{diaz-uriarte2025}, mutation rates, or population size. This simplification is arguably appropriate to start gaining an understanding of the potential use of these methods to identify therapeutic targets and guide interventions (see also \citealp{nichol2015}).

\begin{figure}[tb!]
    \centerline{\includegraphics[width=15.0cm,keepaspectratio]{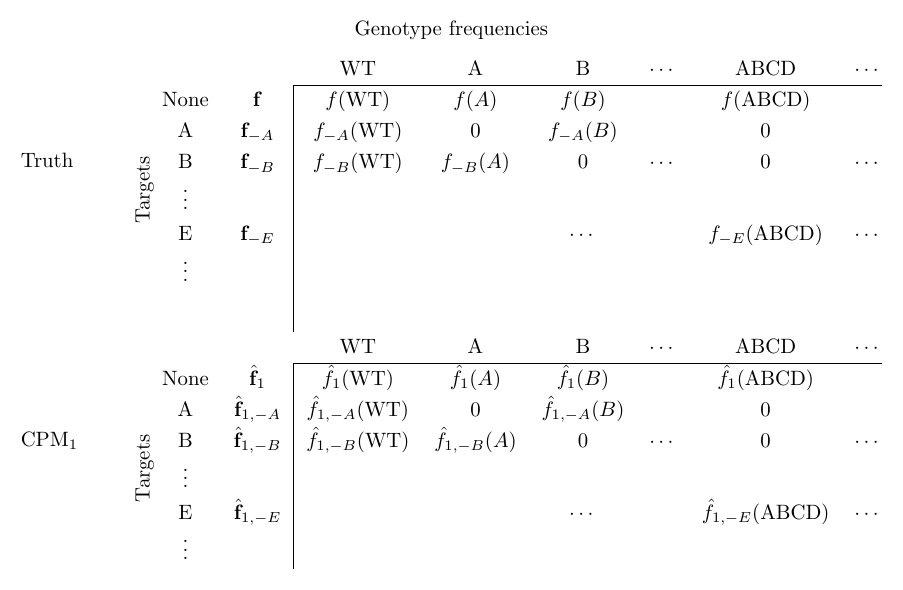}}
    \caption{\textbf{True and predicted hitting probabilities and genotype frequencies under interventions.}  As explained in the text, from the transition rate matrix, \(\mathcal{Q}\), of the true fitness landscape we can obtain the true distribution of genotypes under no intervention (row ``None''), which yields vector \(\mathbf{f}\), with individual entries for each genotype \(f(WT), f(A), \ldots\); we can also obtain the corresponding genotype frequency vectors when intervening on each possible gene (\(\mathbf{f}_{-A}, \mathbf{f}_{-B}, \ldots\)). Similarly, we can obtain the hitting probabilities  under no intervention, \(\mathbf{h}\), and under each possible intervention    (\(\mathbf{h}_{-A}, \mathbf{h}_{-B}, \ldots\)).
      (For simplicity, this table only shows genotype frequencies, but a similar construction exists for hitting probabilities.)
      If using \(\mathcal{Q}\) is not possible, we can use forward genetic simulators to estimate \(\mathbf{f}\) and \(\mathbf{h}\).
For each EvAM \(m\) (for simplicity, only one shown), after fitting the corresponding model, we can obtain the predicted genotype frequencies \(\hat{\mathbf{f}}_m\) and hitting probabilities \(\hat{\mathbf{h}}_m\); after modifying the fitted models to predict the consequences of an intervention (section \ref{sec:cpm-modify-for-intervention}), we obtain the predicted
genotype frequencies (\(\hat{\mathbf{f}}_{m, -A}, \hat{\mathbf{f}}_{m, -B}, \ldots\)) and
hitting probabilities (\(\hat{\mathbf{h}}_{m, -A}, \hat{\mathbf{h}}_{m, -B}, \ldots\)) under each intervention. In the table, some entries are necessarily 0, as shown (e.g., when intervening on \(g = A\), all genotypes with A mutated have frequency, and hitting probability, of 0).}
\label{fig:truth-pred}
\end{figure}

Then, we can obtain the hitting probabilities and distribution of genotypes after intervening on each gene in turn. For each gene \(g\), this is done by first setting the fitness of any genotype that carries a mutation in the target gene \(g\) to 0, and then obtaining the transition rate matrix (and transition matrix of the embedded discrete-chain) that corresponds to this new fitness landscape.
We denote by \(\mathcal{Q}_{-g}\) the true transition rate matrix after intervening on gene \(g\), i.e., after making all genotypes with a mutation in gene \(g\) lethal. %
Using the transition matrix of the embedded discrete-time chain we can obtain the hitting probabilities under intervention (\(\mathbf{h}_{-g}\)); using the transition rate matrix, we can obtain the distribution of genotypes (\(\mathbf{f}_{-g}\)), according to a prespecified sampling scheme%
; these constitute the ground truths of the hitting probabilities and distribution of genotypes that correspond to targeting gene \(g\); see Fig.~\ref{fig:truth-pred}.

Alternatively, if we are not willing to assume SSWM, we can use any forward genetic simulator that operates on fitness landscapes and includes flexible sampling mechanisms, such as OncoSimulR \citep{diaz-uriarte2017} or TumorSim.jl \citep{fontaneda2023}. Using a sensibly large number of simulated runs we can obtain estimates of  \(\mathbf{h}, \mathbf{f}\), as well as \(\mathbf{h}_{-g}, \mathbf{f}_{-g}\) for every gene.

\subsection{Obtaining the predictions of intervention effects from EvAMs}\label{sec:cpm-interv-pred}

The predictions from each EvAM method can be obtained as follows:
\begin{enumerate}

\item From the true \(\mathcal{Q}\) (section \ref{sec:ground-predictions}) (or,  if \(\mathcal{Q}\) is not available for the evolutionary regime considered, using simulations), generate a sample.

\item Use this sample as input for each EvAM method.

\item \label{EvAM_I_g}From the output of each EvAM, obtain a modified EvAM after intervening on (making lethal the mutated allele of) gene \(g\). Let $M_m$ denote the fitted model from method $m$; $M_{m, -g}$ is the modification of \(M_m\) that results from the killing intervention on \(g\) (i.e., after the intervention that makes a mutant in gene \(g\) a lethal mutation). How we obtain \(M_{m,-g}\) is explained in section \qref{sec:cpm-modify-for-intervention}.

\item \label{Pred_EvAM_I_g}From \(M_{m, -g}\), use the standard procedures for each method (Appendix, \qpref{sec:genot_predictions_cpms_error}) to obtain the predicted hitting probabilities, \(\hat{\mathbf{h}}_{m,-g}\), and the predicted distribution of genotypes, \(\hat{\mathbf{f}}_{m,-g}\), after targeting gene \(g\). For example, in Fig.~\ref{fig:truth-pred} this yields each of the rows after ``None'' in panel ``CPM\(_1\)''.

  Hitting probabilities do not depend on sampling time; they are obtained from the transition matrix of the embedded discrete-time chain for CBN, H-ESBCN, and MHN, and from the transition matrix of HyperHMM; for OT and OncoBN we use a heuristic procedure similar to \citet{diaz-uriarte2019a} ---Appendix, \qref{sec:genot_predictions_cpms_error}.

\end{enumerate}

\subsection{Ranking of interventions}\label{sec:ranking-interv}

For each of these three objectives we could focus on the causal effect defined as the difference in one statistic (e.g., the hitting probability of some genotype), between the non-intervened regime and the intervention in each gene ---a framing in the spirit of \citeauthor{hernan2020}'s  ``causal effects in populations, that is, numerical quantities that measure changes in the distribution of an outcome under different interventions'' \citeyearpar[][p.\ v]{hernan2020}.  But, if the objective is to prioritize intervention candidates, it is more useful to reframe the problem as one of ranking interventions.

For O\_mut and O\_WT the most promising EvAMs would be those that rank the interventions, with respect to those two measures, in the same way as the truth. For O\_genotype, the focus is not to compute the rank correlation between
\(\mathbf{h}_{-g}\) and \(\hat{\mathbf{h}}_{m,-g}\);
this would emphasize the agreement between the hitting probabilities of all the genotypes for each particular \(g\). Instead, for each genotype whose frequency we might want to increase or decrease, compute the rank correlation between the true and the predicted hitting probability of that genotype under all possible interventions. This is the procedure that directly answers ``what should I target if I want to increase/decrease the hitting probability of this genotype''. Details in Appendix, section \ref{sec:assess_o_genotype}.

\subsection{Predicted genotype frequencies: at what time?}\label{sec:pred-genot-freqs-time}

True genotype frequencies will likely depend on time for both clinical trial data (when patients are measured) and simulated data (when samples are obtained from the evolutionary process in section \ref{sec:ground-predictions}). For EvAMs, predicted genotype frequencies can depend on time for some models.

The simplest case is when samples under intervention are obtained under a sampling regime (e.g., exponential with some unknown rate, uniform, etc) that is similar to the sampling regime of the non-intervened processes that generated the training data for the EvAMs.
In this case, for CBN, H-ESBCN, and MHN, the predictions of the intervened-upon models can use the standard ``exponential with rate 1'' model: these models are fitted assuming that the true sampling distribution is an exponential of rate 1, regardless of the true sampling distribution (which would not be known to the method anyway). Therefore, for these models we can obtain the predicted genotype frequencies as per equation (4) in \citet{schill2020}, assuming sampling time is distributed as an exponential of rate 1.
For OncoBN and OT, being untimed models, we can obtain the distribution of genotypes as per their standard expressions (which do not include time), setting the model error to 0. It is not possible for these models to match the time at which genotype predictions are made from EvAMs and the true sampling time.
For HyperHMM and HyperTraPS, the scheme in  \qref{sec:hyperhmm-interv} already incorporates obtaining samples weighted by \(P(V = v)\), the number of steps.

If the ground truth of interventions is obtained under a sampling scheme that differs substantially from the sampling scheme of the training data, the above procedure is not adequate. For CBN, H-ESBCN, and MHN, we can obtain predicted genotype frequencies at arbitrary times from \(Q\) (as \(\pi^t = \pi^0 \exp\{Q t\}\), where \(\exp\) is the matrix exponential, \(\pi^0\) is the row vector of the population composition at time 0, with 1 for WT and 0 for the remaining elements), but it would be necessary to express the \(t\) of the true sampling time (or distribution of times) in the scale of the \(t\) of the EvAMs model (recall that these were fitted assuming sampling with exponential rate 1). For HyperHMM and HyperTraPS an analogous adjustment of the \(P(V = v)\) would be needed to match the ``true steps'' that correspond to the ground truth. No such adjustments seem directly possible for OT and OncoBN. Further work to address adjusting the time-dependent predictions of EvAMs is warranted, but this is independent of the intervention framework in section \ref{sec:cpm-modify-for-intervention}.

\section{Discussion} \label{sec:discussion}

After differentiating between two types of intervention (killing and inactivating) and explaining why a naive approach to intervention leads to incorrect predictions for EvAMs, we have presented a conceptualization of interventions on each of the currently available EvAM methods (OT, OncoBN, CBN, H-ESBCN, MHN, HyperHMM, HyperTraPS). We show both what an intervention represents, and the  equivalent ways of implementing the intervention.  We then provide a protocol for assessing how useful EvAMs can be for ranking interventions, i.e., choosing therapeutic targets. The only prior work that has considered intervention in this context is \citet[][Fig.~3]{schill2024b}, who considered suppressing an event (equivalent to our inactivating intervention) in MHN; our framework generalizes this to all EvAM families and additionally distinguishes it from the killing intervention.

Our framework is therefore relevant for using EvAMs to design interventions \citep{hosseini2019a} and, more generally, for making evolutionary predictions with the aim of driving evolution towards, or away from, specific directions \citep{wortel2023}. In fact, our work focuses on EvAMs, but the general procedure is not specific to cancer or EvAMs: we take a class of fitted computational models, interpret them as structural causal models, and derive principled intervention predictions using Pearl's \citep{pearl_causality_2009} framework, with a careful model-by-model analysis of where modularity is and isn't justified. This generalizes beyond EvAMs, and is a template for any domain where fitted models can encode causal structure.

As well, the entities that the model applies to, as emphasized in  \citet{diaz-uriarte2025}, are essential to the conceptualization of some of the interventions. The individual-level causal DAGs of \cite{frank97-price-causal} and Otsuka and collaborators \citep{otsuka_causal_2016, otsuka2019, edelaar2023}, in those cases where we can apply them to EvAMs, make fitness an explicit node in the causal graph. And this is what enables the distinction between killing interventions and inactivating interventions ---a distinction that cannot be expressed in the DAG-of-restrictions or matrix-of-effects (MHN) or transition matrix (HyperHMM/HyperTraPS) formalisms. This is a consequence of these models being, at their most basic, models of accumulation: they describe what events occur and in what order, not necessarily the fitness mechanism that drives selection. Other interpretations of these models are possible (such as frequency-dependent fitness; see discussion in \citealp{diaz-uriarte2025}), but none of them introduce fitness as an explicit variable on which one can intervene separately.
When certain conditions are satisfied (SSWM ---Appendix, section \ref{sec:dags-to-fitness}), some of these models (CBN, H-ESBCN) can be mapped to a fitness landscape, enabling the use of individual-level causal DAGs where fitness is an explicit variable and the killing/inactivating distinction becomes expressible.

The assessment of modularity, a crucial assumption in the interventionist account of causation \citep{woodward_making_2003, pearl_causality_2009}, is also closely related to the entities being modeled. The individual-level causal DAGs allow us to think of interventions as surgical modifications of the fitness landscape, where we make lethal all genotypes that carry a given mutation, leaving the fitness of the rest of the genotypes untouched. This makes the feasibility of the intervention (to follow the dictum ``no causation without manipulation'' ---\citealp{holland1986, hernan2020}) and the plausibility of the modularity assumption easy to scrutinize. For instance, under other evolutionary scenarios, like frequency-dependent fitness, this same conceptualization and modularity assumption would be questionable. That said, the individual-level causal DAG and the killing/inactivating distinction it supports do not depend on the SSWM assumption: SSWM is required only for the equivalence between fitness-landscape modifications and EvAM parameter manipulations in CBN and H-ESBCN.

This framework can be applied without modification or extended to deal with several other scenarios. Interventions that affect multiple genes are expressible in our framework; for DAG-of-restrictions, matrix of effects of MHN, and transition matrices we can only use the \(do\) operator, but if individual-level causal DAGs are possible, then we can differentiate between killing and inactivating. This opens the door to searching for optimal combinations of targets.

In those scenarios where the individual-level causal DAG holds, we can extend the conditional intervention to any gene with a more general intervention function, \(\sigma_{\!W}\) that takes an extra argument, say \(L \in \{0, A, B, \ldots\}\), so \(L = 0\) denotes no intervention and therefore when \(L = 0\), \(h = w(A, B, C)\); when \(L= A\), \mbox{\(h = w(A, B, C) \cdot \boldsymbol{1}_{\{A=0\}} \)}; when \(L = B\), \(h = w(A, B, C) \cdot \boldsymbol{1}_{\{B=0\}} \); etc. \(L\) is not a parent of the original \(W\), but conditional intervention functions  can have as inputs variables other than the original parents \citep{correa2020a}. The framework of \citep{correa2020a} (see also \citealp{eberhardt2007, korb2004, korb_bayesian_2011}) would also allow us to consider stochastic interventions, resulting in partial inactivation or fitness modification of a fraction of the population. Similarly, we can incorporate (possibly conditional) activation interventions, \(do(A = 1)\), which would enable modeling events detrimental to the progression; we do not develop this here.

As well, interventions that are expressible in the DAG-of-restrictions, matrix of effects of MHN, and transition matrices of HyperHMM/HyperTraPS can be applied without modifications to EvAM models used with cohorts of intra-tumor phylogenetic trees, such as TreeMHN \citep{luo2023} and HyperTraPS-CT \citep{aga2024}, and only with minor modifications to models with reversible transitions \citep{johnston2024-hypermk, johnston2026}.

The minor modifications needed for reversible models concern the implementation of interventions, not the causal interpretation. Reversibility introduces transitions in both directions between states, which produces cycles in the influence-diagram summary. But in the underlying structural causal model ---where events are indexed by time--- the structural equations remain acyclic: each transition at time \(t\) depends only on the state at times \(s < t\). Reversibility allows the population to revisit states; it does not allow events to cause themselves. Thus the concerns raised in \citet{bongers2021} do not apply here: reversibility is a property of the state-transition graph, not of the underlying causal structure.

A more substantive change concerns which outcome measures remain meaningful under reversibility. In irreversible accumulation models, hitting probabilities are informative because they reflect which paths the chain commits to. In a finite reversible chain on a connected state space, by contrast, hitting probabilities are not useful since they are 1 for any accessible genotype given enough time. The natural outcome measures shift to stationary distributions and time-indexed quantities at biologically relevant timescales. Adapting the present ranking-based evaluation protocol to these alternative outcome measures is straightforward in principle (one ranks targets by their effect on the chosen measure rather than on hitting probability), but the choice of outcome measure itself is no longer obvious and may need to be problem-specific.

Finally, the current framework presents several limitations, worth further theoretical work. It is unclear how, or even if, stochastic interventions ---which are readily considered in the individual-level causal DAGs--- are expressible in the DAG-of-restrictions, matrix of effects of MHN, and transition matrices of HyperHMM/HyperTraPS.  Conditional and time-varying treatment strategies that depend on repeatedly monitoring and possibly correcting the course of action based on the state of the system, such as adaptive therapy \citep{hansen2020a, west2023}, would require a more complex causal framework \citep{hernan2020, tsiatis2020b, chakraborty2013}, even if we can use individual-level causal DAGs. Of course, interventions that do not map cleanly to the conditional or the \(do(g = 0)\) operation cannot be represented in the current framework. Furthermore, (severe) lack of modularity, such as an intervention on gene \(g\) affecting the parameters for gene \(h\), would invalidate the framework; assessing the plausibility of this assumption is context-specific.

\section{Acknowledgments}

Álvaro San Martín and Eric Macías Fasio, for literature searches,  and initial discussion. Members of the Stochastic Biology group at the University of Bergen for discussion.  Supported by grants PID2024-156888OB-I00 funded by MICIU /AEI/10.13039/501100011033 / FEDER, EU and PID2019-111256RB-I00 funded by MCIN /AEI/10.13039/501100011033 to RDU. This work was supported by the Trond Mohn Foundation [project HyperEvol under grant agreement No. TMS2021TMT09 to IGJ], through the Centre for Antimicrobial Resistance in Western Norway (CAMRIA) [TMS2020TMT11].

\section{Code availability}

Code available from \Burl{https://github.com/rdiaz02/scm-interv-evams}.

\clearpage
\begin{appendices}

\section{Transition rate matrix from fitness landscapes under SSWM and fitness landscapes from CBN and H-ESBCN}\label{sec:trm_from_fitness_and_evam}

Here we detail how we obtain fitness landscapes from CBN and H-ESBCN such that, under strong selection weak mutation (SSWM), the transition rate matrix from the fitness landscape is identical to the one from the CBN/H-ESBCN models.

\subsection{Selection coefficients: definition and notation}
\label{sec:defin-select-coeff}

As is customary (e.g., \citealp{gillespie1984, orr_population_2002,  weinreich_darwinian_2006}), we use \(s\) for the selection coefficient. Let $s_{x \rightarrow x+i}$ be the selection coefficient for the mutation of gene $i$ on a genotype \(x\) without gene \(i\) mutated. Thus, the fitness of a genotype \(x\) that gains mutation $i$, $W_{x+i}$,  is $(1 + s_{x \rightarrow x+i}) W_x$%
. In other words, $s_{x \rightarrow x+i} = (W_{x+i} - W_x)/W_x$, where $W_{x+i}$ is the fitness of genotype $x$ with the addition of gene $i$ mutated.

We use the notation \(s_{x \rightarrow x+i}\) to make it explicit that the fitness effect of gaining mutation \(i\) can depend on background (i.e., there might not be a single, unique \(s_i\)).
For EvAMs that specify deterministic restrictions using DAGs, we can simplify the notation and write just $s_i$ for the selection coefficient for mutation $i$ when the restrictions for the acquisition of mutation $i$ are satisfied.  When the restrictions are not satisfied, any genotype with mutation $i$ has fitness zero, so we need not consider this transition;  if restrictions are satisfied,  the selection coefficient is the same regardless of genotype background.%
In some descriptions below, we will use \(s_i\) when it is clear, from the context, whether we mean \(s_i\) as in EvAMs that use DAGs,  or the more general $s_{x \rightarrow x+i}$ that allows the effect of a mutation \(i\) to depend on the genotype.

\subsection{From fitness landscape to transition rate matrix}\label{sec:fl-to-trm}

We assume SSWM (\citealp{gillespie1984, orr_population_2002}; see also \citealp{gillespie1994, gillespie1983, weinreich_darwinian_2006, nichol2015, krug2021}): $N s_i \gg 1$ and $N \mu \ll 1$, where $\mu$ is mutation rate and $N$ is population size. %
The $s_i$ are sufficiently small that their probability of fixation is approximately $2 s_i$ (\citealp[][p.~232]{gillespie1994}; \citealp[][p.~1319]{orr_population_2002}). Then, in a haploid population the time to fixation is exponentially distributed with mean  $1/(2 N \mu s_i)$, so the transition rate is \(2 N \mu s_i\) (we can use results from haploids, because we are dealing with clonal, or asexual, reproduction ---e.g., \citealp[][p.~50]{charlesworth_elements_2010}; \citealp[][pp.~185--189]{hamilton_pop_gen}; \citealp[][section 3.3]{Otto2007}).

In addition, we make the stronger assumption that  ``evolution only proceeds uphill in the fitness landscape, i.e., it only evolves by fixing beneficial mutations'' \citep[section 6.2]{krug2021}. Thus, we exclude tunneling and escaping from local fitness maxima \citep{Weinreich2005, weinreich_darwinian_2006,  Weissman2009}, which can occur even under SSWM. Therefore, under our model, the probability of transition to a genotype of smaller fitness is 0 (see also, for a similar approach, \citealp[eq.~3]{nichol2015}). We make this simplifying assumption so that we do not need to consider population size and mutation rates (which affect escape from local fitness maxima and tunneling --- \citealp{Weinreich2005, Weissman2009}) nor the role of bulk sequencing \citep{diaz-uriarte2025}.

\subsubsection{Scaling the transition rate matrix}
\label{sec:scaling-trm}

We can scale the transition rate so that the process is faster or slower, or expressed in different units of time (for example, \citealt{gillespie1984}, shows expressions for the scaling of the transition rates so that time is in units of $N$ generations).  We refer to this as the ``scaled transition rate matrix''.

\subsection{CBN and H-ESBCN: From DAGs of restrictions to fitness landscapes; equivalence of transition rate matrices}\label{sec:dags-to-fitness}

We can generate a fitness landscape from CBN and H-ESBCN such that the fitness landscape is not only representable \citep{diaz-uriarte2018, diaz-uriarte2025}, but also yields a transition rate matrix identical (up to a scaling constant) to that of the original CBN/H-ESBCN model. When the transition rate matrices match this way, then in the limit of very large samples from repeated evolution experiments under SSWM on this fitness landscape, fitting the corresponding CBN/H-ESBCN, and assuming the correct parameters are inferred, the transition rates between genotypes from the parameters of the CBN/H-ESBCN model would match the transition rates between genotypes from the fitness landscape. Furthermore, obtaining identical transition rate matrices from both the CBN/H-ESBCN model and its implied fitness landscape means that interventions performed by directly manipulating the fitness landscape or the CBN/H-ESBCN model should yield identical results as interventions via \(DAG_{\lambda_g = 0}\), \(DAG_{-g}\), and \(Q_{-g}\) (section~\ref{sec:cbn-dag-restr-interv}), as confirmed by our tests (file \texttt{kill-gene-equivalences-TESTS.R}).

For CBN and H-ESBCN, the transition rate \(x \rightarrow x+i\) is equal to \(\lambda_i\) when the restrictions for the acquisition of \(i\) are satisfied. And, under our assumed evolutionary regime, that transition rate is \(\propto s_i\) (section \ref{sec:fl-to-trm}), where \(s_i (= s_{x \rightarrow x+i}) = (W_{x+i} - W_x)/W_x\) (section \ref{sec:defin-select-coeff}).
Thus, starting from the WT genotype (e.g., with fitness 1), we can obtain the fitness of all accessible genotypes (those that can exist under the model) by setting \(s_i = a \lambda_i\), and iteratively computing \(W_{x+i} = W_x (1 + s_i)\).

\begin{figure}[tb!]
  \centerline{\includegraphics[width=15.70cm,keepaspectratio]{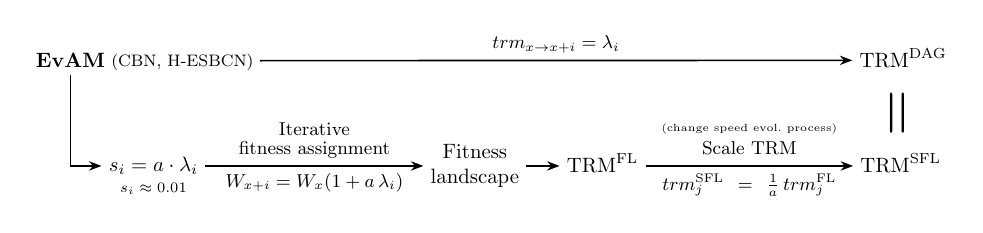}}
  \caption{\textbf{DAGs of restrictions to fitness landscapes: equivalence of transition rate matrices: \(\mathrm{TRM}^{\mathrm{DAG}} = \mathrm{TRM}^{\mathrm{SFL}}\).}\\
\(\mathrm{TRM}^{\mathrm{DAG}}\): transition rate matrix from the DAG of restrictions for CBN and H-ESBCN (\ref{sec:dags-to-fitness}).\\
\(\mathrm{TRM}^{\mathrm{FL}}\): transition rate matrix from the fitness landscape (\ref{sec:fl-to-trm}).\\
\(\mathrm{TRM}^{\mathrm{SFL}}\): scaled transition rate matrix (\ref{sec:scaling-trm}).\\
\(trm^{\mathrm{SFL}}_j, trm^{\mathrm{FL}}_j\): individual entries of \(\mathrm{TRM}^{\mathrm{SFL}}\) and \(\mathrm{TRM}^{\mathrm{FL}}\), respectively.\\
\(s_i, \lambda_i, W_x, W_{x+i}\): sections \ref{sec:defin-select-coeff} and \ref{sec:dags-to-fitness}.\\
\(a = 0.006\) (section \ref{sec:dags-to-fitness} for rationale).
}
\label{fig:evam-fl-trm-scale}
\end{figure}

This will produce a fitness landscape so that the transition rate matrices from CBN and H-ESBCN and the fitness landscape are identical (up to a scaling constant). If we then scale the fitness-landscape derived transition rate matrix (section \ref{sec:scaling-trm}) by multiplying each transition rate by \(1/a\), the transition rate matrix that we obtain directly from the EvAM models (for CBN and H-ESBCN) and the scaled transition rate matrix from the fitness landscape, will be identical (and the individual entries equal to \(\lambda_i\)): Fig.~\ref{fig:evam-fl-trm-scale}.

What is an appropriate value of \(a\) depends on the distribution of the \(\lambda_i\)s.  For example, in MC-CBN (\url{https://github.com/cbg-ethz/MC-CBN}) and EvAM-Tools \citep{diaz-uriarte2022a}, randomly generated CBN models have, by default, \(\lambda \sim \mathcal{U}(1/3, 3)\) so the average $\lambda$  will be close to $1.667$. Thus, a value of $a = 0.006 (\approx 0.01/1.667)$ will generally be adequate to ensure that the \(s_i = a \lambda_i\) are small enough (\(\approx 0.01\)) for SSWM so that the entries of the transition rate matrix are proportional to the \(s_i\).

This correspondence between EvAM models and fitness landscapes is only available for CBN and H-ESBCN.  For EvAMs with stochastic dependencies (MHN, HyperHMM, HyperTraPS), there is no clear mapping between these models and fitness landscapes \citep{diaz-uriarte2025}.
For OT and OncoBN, it is not possible to directly map an OT or OncoBN model to a fitness landscape as OT and OncoBN are untimed models; however, the deterministic dependencies that OT and OncoBN encode are a subset of those captured by CBN and H-ESBCN, so these types of dependencies are mapped to fitness landscapes via CBN/H-ESBCN.

\section{Assessing intervention objectives: a protocol}
\label{sec:assess-interv-object}

Here we provide details on how to assess whether EvAMs can identify promising genes with respect to the three objectives considered in section \ref{sec:quant-interv-effects}.

For all three measures, O\_genotype, O\_mut, and O\_WT, the ground truth is obtained from a fitness landscape in which the targeted gene's manipulation ---setting the fitness of affected genotypes to 0--- has been carried out (this could possibly be done experimentally, though we focus here on the computational manipulation of setting fitness to 0). The predictions of the interventions are then obtained from an EvAM method. Finally, predictions and ground truth are compared. None of these measures require that the EvAM method have a fitness landscape interpretation: they require only that we can obtain hitting probabilities (O\_genotype) or genotype frequencies (O\_mut, O\_WT) from the EvAM. The fitness landscape is needed only for the ground truth, not for the predictions.

\begin{figure}[tb!]
  \centerline{\includegraphics[width=16.0cm,keepaspectratio]{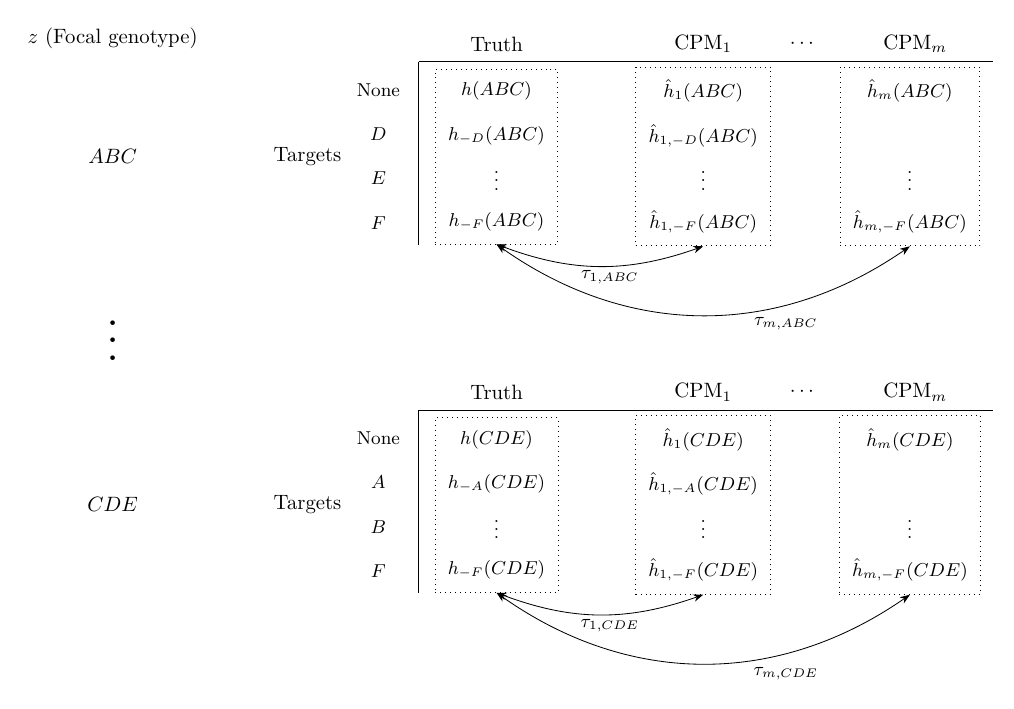}}
  \caption{\textbf{Assessing O\_genotype: Increase/decrease the probability of going through (hitting) a given genotype.} For each method, we can compute Kendall's \(\tau\) correlation between the truth and the predictions of the hitting probability under all possible interventions for every focal genotype (the figure shows two focal genotypes, \(\{A, B, C\}\) and \(\{C, D, E\}\)). For example, in the bottom row, \(h_{-F}(CDE)\) is the true hitting probability of genotype \(\{C, D, E\}\) when we intervene on gene \(F\), and \(\hat{h}_{1, -F}(CDE)\) the predicted hitting probability of genotype \(\{C, D, E\}\) when we intervene on gene \(F\) from CPM\(_1\). For a given method, the value of the statistic is the average rank correlation over all the focal genotypes (so for method \(\mathrm{CPM}_1\) we would average \(\tau_{1,ABC}, \ldots, \tau_{1,CDE}\) ).}
\label{fig:o_genotype}
\end{figure}

\subsection{Increase/decrease the probability of going through (hitting) a given genotype
(O\_genotype)}\label{sec:assess_o_genotype}

We assess whether EvAMs can correctly predict which genes, when targeted, most decrease (or increase) the  probability of hitting (going through) a given genotype, the ``focal genotype''. When considering a particular focal genotype, \(z\), the set of possible target genes cannot include any of the genes mutated in \(z\) since the trivially best approach to have a hitting probability of 0 would be to target any of the mutated genes in \(z\).

We might want to consider, as focal genotypes, all accessible genotypes in the true fitness landscape that have a certain number of mutations. The rationale is to evaluate performance over a large number of genotypes that can capture complex dependencies and are assessed under a range of interventions. For example, in a simulation study involving 9 genes, we might want to focus on genotypes in the true fitness landscape that have between 3 and 5 mutations. This removes from consideration genotypes with only 1 or 2 mutations (too few mutations to reflect complex dependencies), and at the same time all genotypes assessed are evaluated under at least 5 different interventions, so that the correlation coefficient between truth and prediction is computed with at least 5 points (as we will probably want to use the ``no intervention'' as an additional point).

  For a given fitness landscape we can assess this objective as:

  \begin{enumerate}[label*=\arabic*.]

  \item Focal genotypes are all accessible genotypes in the true fitness landscape with a certain number of mutations (e.g., 3 to 5 mutations); we use \(z\) to denote any one of the focal genotypes.\label{o_genotype_trmii0}

\item If under SSWM, from \(\mathcal{Q}\), the true transition rate matrix of the fitness landscape, obtain the transition matrix corresponding to the embedded discrete-time chain; from it, obtain the hitting (or first passage) probabilities (\citealp{norris1997, privault2018}), \(\mathbf{h}\), starting from the WT state  (hitting probabilities can be readily obtained using, for example, the \texttt{markovchain} R package --- \citealp{spedicato2017}). Thus, \(h(z)\), the entry of \(\mathbf{h}\) corresponding to a focal genotype \(z\), is the hitting probability of genotype \(z\) (and we make no difference between transient and absorbing genotypes, so the hitting probability of an absorbing genotype is also its absorption probability). Alternatively, if \(\mathcal{Q}\) is not available for the evolutionary regime considered, use simulations to estimate \(\mathbf{h}\).\label{o_genotype_trmii01}

\item For every gene \(g\), obtain \(\mathcal{Q}_{-g}\), the transition rate matrix when intervening on gene \(g\) and, from it, the transition matrix corresponding to the embedded discrete-time chain. From it, obtain the hitting probability of genoypes under the intervention, \(\mathbf{h}_{-g}\). \(h_{-g}(z)\) is the hitting probability of genotype \(z\) when intervening on gene \(g\). (If \(\mathcal{Q}_{-g}\) is not available for the evolutionary regime considered, use simulations to estimate \(\mathbf{h}_{-g}\).)

  \label{o_genotype_trmii}

  \item For each EvAM, \(M_m\):
    \begin{enumerate}[label*=\arabic*.]
    \item From \(M_m\), obtain the predicted hitting probability of genotypes under no intervention \(\hat{\mathbf{h}}_{m}\). \(\hat{h}_{m}(z)\) is the hitting probability, for method \(m\), of genotype \(z\) under no intervention.\label{o_genotype_cpmii0}

    \item Modify the model results under each intervention to obtain \(M_{m, -g}\) and, from it, the predicted hitting probability of genotypes \(\hat{\mathbf{h}}_{m,-g}\).  \(\hat{h}_{m, -g}(z)\) is the predicted hitting probability, under model \(m\), of genotype \(z\) when intervening on gene \(g\).     \label{o_genotype_cpmii}

    \item  For each focal genotype, \(z\):
      \begin{enumerate}[label*=\arabic*.]
      \item For each EvAM, compute the rank correlation between the vectors\\ \((h(z), h_{-A}(z), h_{-B}(z), \ldots )\) and \((\hat{h}_{m}(z), \hat{h}_{m, -A}(z), \hat{h}_{m, -B}(z), \ldots )\).

        In other words, compute the rank correlation between the true (from steps \ref{o_genotype_trmii0} and \ref{o_genotype_trmii}) and the predicted hitting probability of \(z\) (from steps \ref{o_genotype_cpmii0} and \ref{o_genotype_cpmii}) under all possible interventions (i.e., those that target a gene \(g\), where \(g\) is not one of the genes mutated in \(z\)); include also the ``no intervention'' as a data point (it is conceivable that some intervention could be worse than doing nothing).

      Here, and elsewhere, we suggest measuring correlation using Kendall's \(\tau\) (not Spearman's rank correlation), to emphasize choosing one intervention vs.\ another (i.e., pair agreement) rather than overall ordering agreement. (Differences in results from different correlation coefficients are, however, unlikely to be relevant.)

   Note that what we would compute is different from the rank correlation between \(\mathbf{h}_{-g}\) and \(\hat{\mathbf{h}}_{m,-g}\); this would emphasize the agreement between the hitting probabilities of all the genotypes for each particular \(g\). This is not as useful as the measure proposed here, which directly answers ``what should I target if I want to increase/decrease the hitting probability of genotype \(z\)''.

      \end{enumerate}
      \item Return the average rank correlation coefficient over all \(z\).
    \end{enumerate}

  \end{enumerate}

\subsection{Decrease overall frequency of mutations or alterations: O\_mut}\label{assess_o_mut}

For a given fitness landscape we can assess this objective as:

\begin{enumerate}[label*=\arabic*.]

\item From \(\mathcal{Q}\), the true transition rate matrix of the fitness landscape, obtain \(\mathbf{f}\), the distribution of genotypes, using the same time distribution as used when generating the samples (section \ref{sec:ground-predictions}). (As above, if  \(\mathcal{Q}\) is not available, we can instead use forward genetic simulations to estimate \(\mathbf{f}\).)

  From \(\mathbf{f}\), obtain the true population mean number of mutations as \(D = \sum_z n(z) f(z)\), where \(n(z)\) is the number of mutations of genotype \(z\) and \(f(z)\) is as defined above.

  \item For every \(g\), from \(\mathcal{Q}\) (or using forward genetic simulations on the modified fitness landscapes), obtain  \(\mathbf{f}_{-g}\), and from it the true mean number of mutations under intervention on \(g\): \(D_{-g} = \sum_z n(z) f_{-g}(z)\).

  \item For each EvAM, \(M_m\):
    \begin{enumerate}[label*=\arabic*.]

    \item From \(M_m\), obtain the predicted distribution of genotypes under no intervention \(\hat{\mathbf{f}}_{m}\). \(\hat{f}_{m}(z)\) is the predicted frequency, for method \(m\), of genotype \(z\) under no intervention. From  \(\hat{\mathbf{f}}_{m}\) obtain the predicted mean number of driver mutations, \(\hat{D}_m = \sum_z n(z) \hat{f}_{m}(z)\).
    \item From \(M_{m, -g}\) obtain \(\hat{\mathbf{f}}_{m, -g}\) and from this, the predicted mean number of driver mutations, \(\hat{D}_{m,-g} = \sum_z n(z) \hat{f}_{m, -g}(z)\).

    \item Compute the rank correlation between the vectors\\ \((D, D_{-A}, D_{-B}, \ldots )\) and \((\hat{D}_{m}, \hat{D}_{m, -A}, \hat{D}_{m, -B}, \ldots )\).

      In other words, compute the rank correlation between the true and predicted mean number of mutations under all possible interventions; include also the ``no intervention'' as a data point.
    \end{enumerate}
  \end{enumerate}

  As we are measuring the correlation between the true and predicted effects of changes, EvAMs that perform well here could also be used to choose interventions that increase the overall frequency of mutations.

\subsection{Increase frequency of the non-mutated, starting genotype, WT: O\_WT}\label{assess_o_wt}

Very similar to \ref{assess_o_mut}. For a given fitness landscape we can assess this objective as:

   \begin{enumerate}[label*=\arabic*.]
   \item Obtain \(f(WT)\), i.e., \(f(z)\) when \(z\) is the WT genotype.
   \item For every \(g\), obtain \(f_{-g}(WT)\), the frequency of the WT genotype when intervening on gene \(g\).

  \item For each EvAM, \(M_m\):
    \begin{enumerate}[label*=\arabic*.]
    \item Obtain \(\hat{f}_{m}(WT)\) (predicted frequency of WT for model \(m\)).
    \item For every \(g\), obtain \(\hat{f}_{m,-g}(WT)\) (predicted frequency of WT for model \(m\) under intervention \(g\)). \label{o_wt_cpmii}

  \item For each EvAM, compute the rank correlation between the vectors \\ \((f(WT), f_{-A}(WT), f_{-B}(WT), \ldots )\) and \((\hat{f}(WT), \hat{f}_{-A}(WT), \hat{f}_{-B}(WT), \ldots )\).

  \end{enumerate}
\end{enumerate}

\begin{figure}[tb!]
   \centerline{ \includegraphics[width=14.0cm,keepaspectratio]{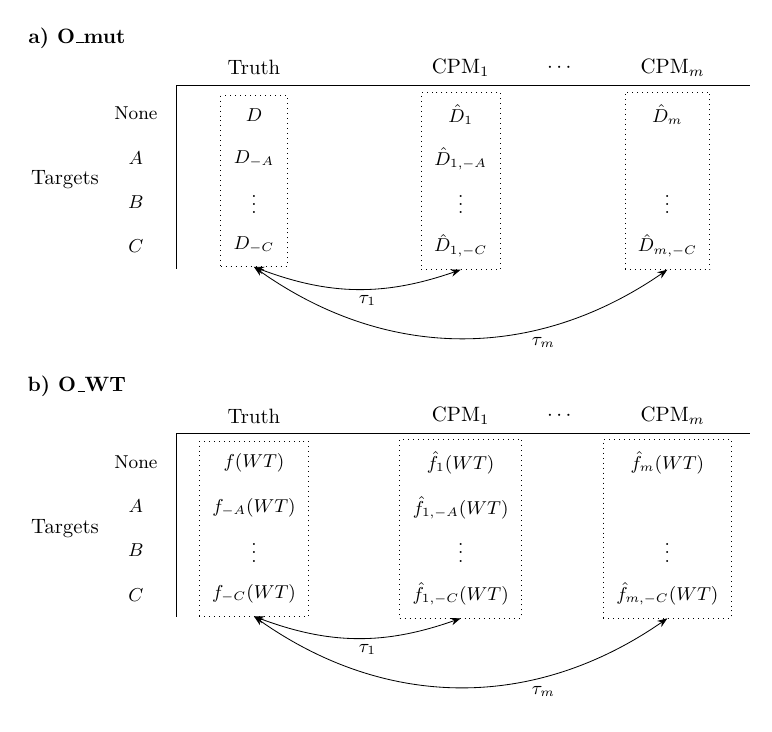}}
 \caption{\textbf{Assessing O\_mut (decreasing overall frequency of mutations) and O\_WT (increasing the frequency of the WT genotype)}. See text for meaning of symbols. For each CPM, the value of the statistic is the rank correlation between predictions and truth.}
\label{fig:o_mut_o_wt}
\end{figure}

\section{Interventions: code and algorithmic details}\label{sec:appendix-interv-details}

\subsection{Interventions as modification of the DAGs: details and algorithm}\label{sec:cpm_interv_dag}

OT, OncoBN, CBN, and H-ESBCN specify the EvAM model using DAGs of restrictions. Our default procedure for targeting gene \(g\) involves eliminating, from the DAG that specifies the model, the node of the targeted gene and any other descendant nodes that strictly depend on the targeted gene. We denote this as \(DAG_{-g}\). Specifically:

\begin{itemize}
\item OT: the models are trees, and thus a node depends on only one ancestor. If gene \(h\) depends on targeted gene \(g\), making \(g\) lethal precludes \(h\) and all of its descendants from accumulating. Thus, for OT, we remove from the DAG the targeted gene \(g\) and all of its descendants.

  \item CBN: the models are DAGs, with AND conjunction. If a gene \(h\) depends on the targeted \(g\) and some other genes \(o_1, o_2, \ldots\), making \(g\) lethal means \(h\) and all of its descendants must be removed from the DAG too. Thus, for CBN, we remove from the DAG the targeted gene \(g\) and all of its descendants.

  \item For OncoBN, under the disjunctive model, a gene \(h\) can depend, with an OR relationship, on \(g\), the targeted gene, and other genes \(o_1, o_2, \ldots\); in other words, \(h\) should be removed if and only if it depends only on \(g\). Thus, for OncoBN under the disjunctive model we remove from the DAG the targeted gene \(g\) and all of its descendants that depend only on \(g\). For OncoBN under the conjunctive model we operate as for CBN.

    \item For H-ESBCN, a gene  \(h\) can depend on \(g\), the targeted gene, and other genes \(o_1, o_2, \ldots\). The dependency can be conjunctive (AND) as for CBN, disjunctive (OR) as for OncoBN, and exclusive OR (XOR); the type of dependency of a gene on its immediate ancestors is the same for all ancestors. Thus, for H-ESBCN, we  remove the targeted gene \(g\) and its descendants as dictated by whether the dependency is AND (similar to what we do with CBN) or OR/XOR (similar to what we do with OncoBN).

\end{itemize}

In all the cases above, when we target gene \(g\), a gene \(h\) should be removed from the DAG if and only if \(h\) appears in accessible genotypes only with gene \(g\). This leads to a simple algorithm for removing genes from the DAG that does not require traversing the DAG to assess exclusive dependencies, and that is easy to implement if the DAG is stored as a collection of pairs of edges \(From, To\):

\begin{enumerate}
\item Identify the genotypes that become non-viable after the intervention: all those that have \(g\).
\item Find all genes, \(h_1, h_2, \ldots\) that appear in formerly viable genotypes only with \(g\).
\item Remove, from the DAG, any edges that involve \(g, h_1, h_2, \ldots\) (i.e., any edges that involve \(g, h_1, h_2, \ldots\) as origin or destination).
\end{enumerate}

\code{The above algorithm is in function \texttt{kill\_gene\_DAG} (called from \texttt{kill\_gene}) in file \texttt{kill-gene-and-output-from-cpm.R}.}

\code{Function \texttt{intervene\_cpm\_every\_gene}, in file \texttt{intervention.R}, is the main intervention function, which calls \texttt{kill\_gene} and obtains predicted genotype frequencies and hitting probabilities (function \texttt{get\_genotype\_freqs\_cpm}).}

\code{Function  \texttt{kill\_gene\_DAG\_param\_0} called by \texttt{kill\_gene\_by\_params\_to\_0} kills by setting the corresponding parameter (\(\lambda_i, \theta_i, \pi_{pa(i), i}\), for CBN/H-ESBCN, OncoBN, and OT, respectively) to 0; tests showing equivalence of setting the parameter and removing from DAG in file \texttt{kill-gene-equivalences-TESTS.R}.}

\subsection{MHN: Interventions as modifications of the \(\Theta\) matrix and the transition rate matrix}\label{sec:mhn_interv_details}

As explained in \qref{sec:mhn_graph-interv}, we intervene on a gene \(g\), \(do(g = 0)\), by setting \(\Theta_{g,g} = 0\). The net effect of  this is setting   \(Q_{x \rightarrow x+g}\), the transition rate from  genotype \(x\) to genotype \(x\) plus a mutation in gene \(g\), to 0.
To prevent having non-zero transition rates from genotypes that are not accessible, we also set \(\Theta_{i, g} = 0 \), where \(i\) are each of the remaining genes, to ensure that all transitions out of a genotype with \(g\) mutated are 0. (Setting \(\Theta_{g, i} = 0 \) can also be done, but is unnecessary if we have set  \(\Theta_{g, g} = 0\)).

Setting \(\Theta_{g,g} = 0\) amounts to eliminating \(g\) from the \(\Theta\) matrix. This is what we use routinely. \code{Function \texttt{kill\_gene\_MHN} (called from \texttt{kill\_gene}) in file\\ \texttt{kill-gene-and-output-from-cpm.R}.}

\code{Function \texttt{kill\_gene\_MHN\_theta\_minus\_Inf} called by\\ \texttt{kill\_gene\_by\_params\_to\_0}. The code sets \(\theta = -\infty\), as \(\theta\), not \(\Theta\), is the parameter directly used in our code.}

\code{Tests showing equivalence of the two procedures in file\\ \texttt{kill-gene-equivalences-TESTS.R}.}

\subsection{CBN, H-ESBCN, MHN: equivalence of setting parameters to 0 and removing entries from the transition rate matrix}\label{sec:appendix-equiv-remove-trm-param-0}

\code{Tests showing the equivalence of: a) setting the corresponding parameter (\(\Theta_{g,g}, \lambda_g\)) to 0; b) removing, from the transition rate matrices, all genotypes with gene \(g\) mutated (and then computing the distributions of genotypes under sampling with the time distributed as an exponential with rate 1),  in file \texttt{kill-gene-equivalences-TESTS.R}.}

\subsection{HyperHMM: Interventions via creation of a self-transition in the transition probability matrix}\label{sec:hyperhmm-details}

The algorithm has been explained in full in \qref{sec:hyperhmm-interv}.

\code{Function \texttt{kill\_gene\_HyperHMM} in file \texttt{kill-gene-and-output-from-cpm.R}. Tests in file \texttt{tests/kill-HyperHMM-TESTS.R}.}

\subsection{Interventions on EvAMs as modifications of the fitness landscape}\label{sec:appendix-interv-flandscape}

\code{Function \texttt{intervene\_fitness\_landscape\_every\_gene}, in file \texttt{intervention.R} returns the results of intervening on every gene via modification of the fitness landscape. This function calls \texttt{ kill\_gene\_fitness\_landscape} which is the one that does the actual modification of the fitness landscape for all genotypes affected by killing gene \(g\). }

\code{Equivalence of killing via fitness landscapes modification and DAG of restrictions intervention is shown in file \texttt{kill-gene-equivalences-TESTS.R}, section ``CBN and H-ESBCN: intervene by modifying the fitness landscape identical to DAG intervention''.}

\subsection{Genotype predictions and hitting probabilities from EvAMs and error model}\label{sec:genot_predictions_cpms_error}

For CBN, H-ESBCN, MHN we can obtain the genotype frequency predictions from the transition rate matrix. This transition rate matrix is obtained directly from the expressions of MHN \citep{schill2020}, and can be obtained from the DAG of restrictions and the values of the set of \(\lambda\) parameters for CBN and H-ESBCN \citep{montazeri_large-scale_2016,angaroni2021} (see also \citealp{diaz-uriarte2025}). From the transition rate matrix, the predicted genotype frequency is obtained as per equation (4) in \citet{schill2020}, assuming sampling time is distributed as an exponential of rate 1.

For OT and OncoBN, predicted genotype frequencies are a function of both the conditional probabilities of the model and deviations from the models. For OT the error model includes possibly different false positive and false negative observation errors and deviations from the model; the observation error is reflected in both the estimated false positive \(\epsilon_+\) and the estimated false negative  \(\epsilon_-\) errors, but the true deviations from the model are reflected only in \(\epsilon_+\) (see \citealp{Szabo2002,Szabo2008}, and Supplementary Material to \citealp{diaz-uriarte2022a}). OncoBN has a different error model from OT, with a ``spontaneous activation model'' that allows child mutations to accumulate even if the parents in the DAG have not been mutated \citep{nicol2021} (see also Supplementary Material to \citealp{diaz-uriarte2022a}, section 5.2). So as to use a similar procedure when generating predicted genotype frequencies as the one for CBN/H-ESBCN/MHN, we can set the error terms ($\epsilon_+$ ---\texttt{epos}--- and  $\epsilon_-$ ---\texttt{eneg}--- for OT and $\epsilon$ for OncoBN) to 0 (see also discussion in sections 3.2 and 5 in the Supplementary Material to \citealp{diaz-uriarte2022a}). \code{In function \texttt{get\_full\_output}, \texttt{epos} is set to 0 by default; see file \texttt{kill-gene-and-output-from-cpm.R}. \texttt{get\_full\_output} is called from \texttt{get\_genotype\_freqs\_cpm}. %
}

For CBN, H-ESBCN, and MHN hitting probabilities can be obtained from the  transition matrix of the embedded discrete-time chain. For HyperHMM/HyperTraPS from the transition matrix of HyperHMM/HyperTraPS. For OT and OncoBN we can use the same procedure as used in \citet{diaz-uriarte2019a} (section 3.1 in Supplementary file S4); thus, we construct a matrix where the ``genotype transition rate'' entries (with 0 in the diagonal) are the conditional probabilities (\(\pi_{pa(i), i}, \theta_i\), for OT and OncoBN, respectively) and then row-scale that matrix so that entries sum to 1 across rows. This is the ``transition matrix'' used to obtain the hitting probabilities. This is a heuristic procedure, because OT and OncoBN are untimed models that provide no transition matrices (nor transition rate matrices).

\subsection{Interventions: ground truth from fitness landscapes.}\label{sec:appendix-flandscape-ground-truth}

\code{Function \texttt{intervene\_fitness\_landscape\_every\_gene}, from \texttt{intervention.R}, which calls \texttt{kill\_gene\_fitness\_landscape} from file \texttt{kill-gene-and-output-from-cpm.R}}

\end{appendices}

\end{document}